\pdfoutput=1

\documentclass[twocolumn]{emulateapj}

\usepackage{amsmath}
\usepackage[figuresleft]{rotating}
\usepackage{float}
\usepackage{graphicx}
\usepackage{color}
\usepackage[colorlinks]{hyperref}
\hypersetup{
    colorlinks,	
    citecolor=blue,
}

\usepackage{multirow}

\newcommand{\srcg}{GX~339$-$4}
\newcommand{\srce}{EXO~1846$-$031}

\newcommand{\nicer}{\textit{NICER}}
\newcommand{\hxmt}{HXMT}

\newcommand{\be}{\begin{eqnarray}}
\newcommand{\ee}{\end{eqnarray}}

\shorttitle{Evolution of QPOs in \srcg\ and \srce }
\shortauthors{Zhang et al.}

\begin{document}

\title{Evolution of QPOs in \srcg\ and \srce\ with \textit{Insight}-HXMT and \nicer }

\author{Zuobin~Zhang\altaffilmark{1}, Honghui~Liu\altaffilmark{1}, 
Divya~Rawat\altaffilmark{2},
Cosimo~Bambi\altaffilmark{1,3,\dag},
Ranjeev~Misra\altaffilmark{4},
Pengju~Wang\altaffilmark{5,6},
Long~Ji\altaffilmark{7},
Shu~Zhang\altaffilmark{5,6} and
Shuangnan~Zhang\altaffilmark{5,6}
}

\altaffiltext{1}{Center for Astronomy and Astrophysics, Center for Field Theory and Particle Physics, and Department of Physics,\\
Fudan University, Shanghai 200438, China. \email[\dag E-mail: ]{bambi@fudan.edu.cn}}
\altaffiltext{2}{Observatoire Astronomique de Strasbourg, Universit\'e de Strasbourg, CNRS, F-67000 Strasbourg, France}
\altaffiltext{3}{School of Natural Sciences and Humanities, New Uzbekistan University, Tashkent 100007, Uzbekistan}
\altaffiltext{4}{Inter-University Center for Astronomy and Astrophysics, Ganeshkhind, Pune 411007, India}
\altaffiltext{5}{University of Chinese Academy of Sciences, Chinese Academy of Sciences, Beijing 100049, China}
\altaffiltext{6}{Key Laboratory of Particle Astrophysics, Institute of High Energy Physics, Chinese Academy of Sciences, Beijing 100049, China}
\altaffiltext{7}{School of Physics and Astronomy, Sun Yat-Sen University, Zhuhai 519082, China}


\begin{abstract}
We conduct a spectral and timing analysis of \srcg\ and \srce\ with the aim of studying the evolution of type-C QPOs with spectral parameters. The high cadence data from \textit{Insight}-HXMT and \nicer\ allow us to track the evolution of QPOs and spectra simultaneously. Type-C QPOs appear at the end of low-hard state and/or hard- intermediate state. Our results reveal that the QPO frequency is closely related to the inner disk radius and mass accretion rate in the two sources. This correlation aligns well with the dynamic frequency model of a truncated disk.
\end{abstract}



\section{Introduction}

Quasi-periodic oscillations (QPOs) refer to narrow peak structures in the power density spectrum (PDS), commonly observed in X-ray binaries (XRBs) \citep{van2005}. In black hole systems, QPOs are mainly classified into low-frequency QPOs (LFQPOs, centroid frequency $0.1-30$~Hz) and high frequency QPOs (HFQPOs, centroid frequency $\geq 60$~Hz) \citep{Belloni2010}. \citet{Samimi1979} reported the ‘sporadic quasi-periodic behaviour’ in the light curve of \srcg, and \citet{Motch1983} reported the first rigorous detection of QPOs for the same source. It was immediately recognized that QPOs would have been a powerful tool to study the accretion process around black holes. 

Over the last forty years, especially after the launch of Rossi X-ray Timing Explorer (\textit{RXTE}), we have accumulated a lot of knowledge about QPOs. Using a broken power-law to fit the broadband noise in PDS, and a Lorentz function with the centroid frequency of $f_{\rm QPO}$ to fit the LFQPOs, \citet{Wijnands1999} found a significant positive correlation between the break frequency $f_{\rm b}$ and the frequency of the LFQPOs $f_{\rm QPO}$. \citet{Psaltis1999} reported a significant positive correlation between the frequency of LFQPOs and the frequency of the broadband noise (or HFQPOs) in low-mass XRBs (LMXBs), including black holes and neutron stars.  

We have observed LFQPOs in most black hole XRBs and realized that LFQPOs can be divided into three types: type-A, -B, and -C QPOs, based on quality factor, noise type, fractional rms, and phase delay (e.g., \citealt{Wijnands1999}; \citealt{Sobczak2000}; \citealt{Casella2005}; \citealt{Motta2011}). Different types of QPOs occupy distinct regions on the hardness-intensity diagram, as well as significantly distribute in different areas on the centroid frequency versus rms diagram (e.g., \citealt{Motta2011}). The phenomenon of rapid transition between different types of QPOs has been found in some sources, and the time scale of this phenomenon can be very short (10~s) (e.g., \citealt{Bogensberger2020}). In this work, we only focus on the type-C QPOs.

Type-C QPOs appear in the early stage of the outburst, particularly at the hard-intermediate state and at the end of low-hard state. The centroid frequency varies from a few mHz to $\sim 10$~Hz, and is tightly correlated with the spectral state. \citet{Vignarca2003} reported a positive correlation between the centroid frequency and the photon index $\Gamma$. \citet{Motta2011} found the centroid frequency of type-C QPOs significantly correlate with the corona flux and disk flux. The dependence of the QPO frequency on photon energy was illustrated by \citet{Qu2010}.

In addition to the phenomenological study of QPOs, many studies has been done on the theoretical explanation behind them. Most theoretical models explain the QPO phenomenon through the following two different mechanisms: instabilities of the corona-disk system (e.g., \citealt{Titarchuk2004}; \citealt{Mastichiadis2022}; \citealt{Varniere2012}) or the geometrical effects of general relativity (e.g., \citealt{Stella1998}; \citealt{Ingram2009}). \citet{Titarchuk2004} introduced a transition layer in corona-disk system to explain the QPO phenomenon in XRBs. The disk-corona natural frequency model was proposed by \citet{Mastichiadis2022}, and they argued that type-C QPOs arise from the interaction of the hot corona with the cold accretion disk. \citet{Varniere2012} suggested that LFQPOs could result from the relativistic accretion-ejection instability (AEI). 

\citet{Ingram2009} interpreted the QPOs in LMXBs as Lense–Thirring precession of the hot flow inside a truncated disk. Also in the frame of Lense–Thirring precession, \citet{Stella1998} proposed LFQPOs is due to precession of the innermost region of accretion disk. In recent years, more and more models and observational evidences have been reported (e.g., \citealt{Karpouzas2020}; \citealt{Garcia2021}; \citealt{Bellavita2022}; \citealt{Mendez2022}; \citealt{Peirano2023}). However, a unified model that can explain all QPO behavior has not been found yet.

Recently, \citet{Misra2020} identified the QPO frequency of GRS~1915+105 as the relativistic dynamic frequency of a truncated accretion disk with \textit{AstroSat} observation data. The authors found a strong correlation between the QPO frequency divided by the accretion rate and the inner disk radius. The correlation is consistent with the prediction of the dynamic frequency model under the assumption of a standard relativistic accretion disk \citep{Novikov1973}. \citet{Liu2021} extended the relation to cover a wider range of variations and confirmed the high spin nature of the black hole in GRS~1915+105 with the data from \textit{Insight}-HXMT (dubbed \hxmt; \citealt{Zhang2014}). We note that GRS~1915+105 is a persistent source with particular properties \citep{Belloni2000}. We would like to test if this relation holds for other sources different from GRS~1915+105, and we notice that there are two appropriate sources, \srcg\ and \srce, in the archive.

The XRBs transient \srcg\ is a typical LMXBs discovered in 1973 \citep{Markert1973}. It undergoes bright outburst every few years, and all four X-ray states typically seen in XRBs have been detected in this system (e.g., \citealt{Miyamoto1995}; \citealt{Homan2005}; \citealt{Plant2014}). \srcg\ is located at $8–12$~kpc with a black hole mass of $4–11$~M$_{\bigodot}$ \citep{Zdziarski2019}. Strong relativistic reflection signatures have been found in this source in both the hard and soft states (e.g., \citealt{Garcia2015}; \citealt{Miller2004}; \citealt{Liu2022}; \citealt{Liu2022_2}). Previous studies have indicated that the black hole in \srcg\ has a very high spin ($a_* \sim 0.95$, \citealt{Garcia2015}; \citealt{Parker2016}). The inclination angle of the accretion disk should have an intermediate value (\citealt{Furst2015}; \citealt{Parker2016}). \citet{Motta2011} systematically studied the properties of QPOs of \srcg\ and suggested that the centriod frequency of QPOs (including type-C QPOs) correlate with the disk flux.

\srce\ was discovered by the European X-ray Observatory Satellite (\textit{EXOSAT}) when it went into outburst in April 1985, and then it was considered an LMXBs (\citealt{Parmar1993}; \citealt{Draghis2020}). \textit{CGRO}/BATSE detected a second outburst in 1994 \citep{zhang1994}. Subsequently, the source remained in a quiescent state for 25 years. In 2019, \srce\ experienced a new outburst, monitored by X-ray missions (e.g., \textit{MAXI}/GSC; \hxmt; \textit{NuSTAR}) and radio telescopes (e.g., \textit{MeerKAT}; \textit{AMI-LA}). A high-quality observation was conducted by \textit{NuSTAR} on August 3, 2019, with a 22.2 ks exposure time. \citet{Draghis2020} reported strong relativistic reflection features with the sensitive \textit{NuSTAR} spectra, concluding that the source is a black hole with a nearly maximal spin parameter ($a_* = 0.997$) at a disk inclination of $\theta = 73^\circ$. \srce\ is located at $2.4–7.5$~kpc, according to the previous studies on X-ray and radio data (\citealt{Parmar1993}; \citealt{Williams2022}), with a black hole mass of $\sim 9$~M$_{\bigodot}$ (\citealt{Draghis2020}; \citealt{Williams2022}). \citet{Liuhx2021} reported the observational results from a detailed timing analysis of \srce\ 2019 outburst using the data from \hxmt\ and \nicer. 

In this work, we focus on the latest \hxmt\ and \nicer\ observations of \srcg\ and \srce, and present a detailed temporal and spectral analysis. The paper is organized as follows. Sec.~\ref{obs} presents the observational data reduction. The spectral timing analysis is reported in Sec.~\ref{analysis}. We discuss the results and report our conclusions in Sec.~\ref{discuss} and Sec.~\ref{conclusion}, respectively.



\begin{figure*}[]
    \centering
    \includegraphics[width=0.9\linewidth]{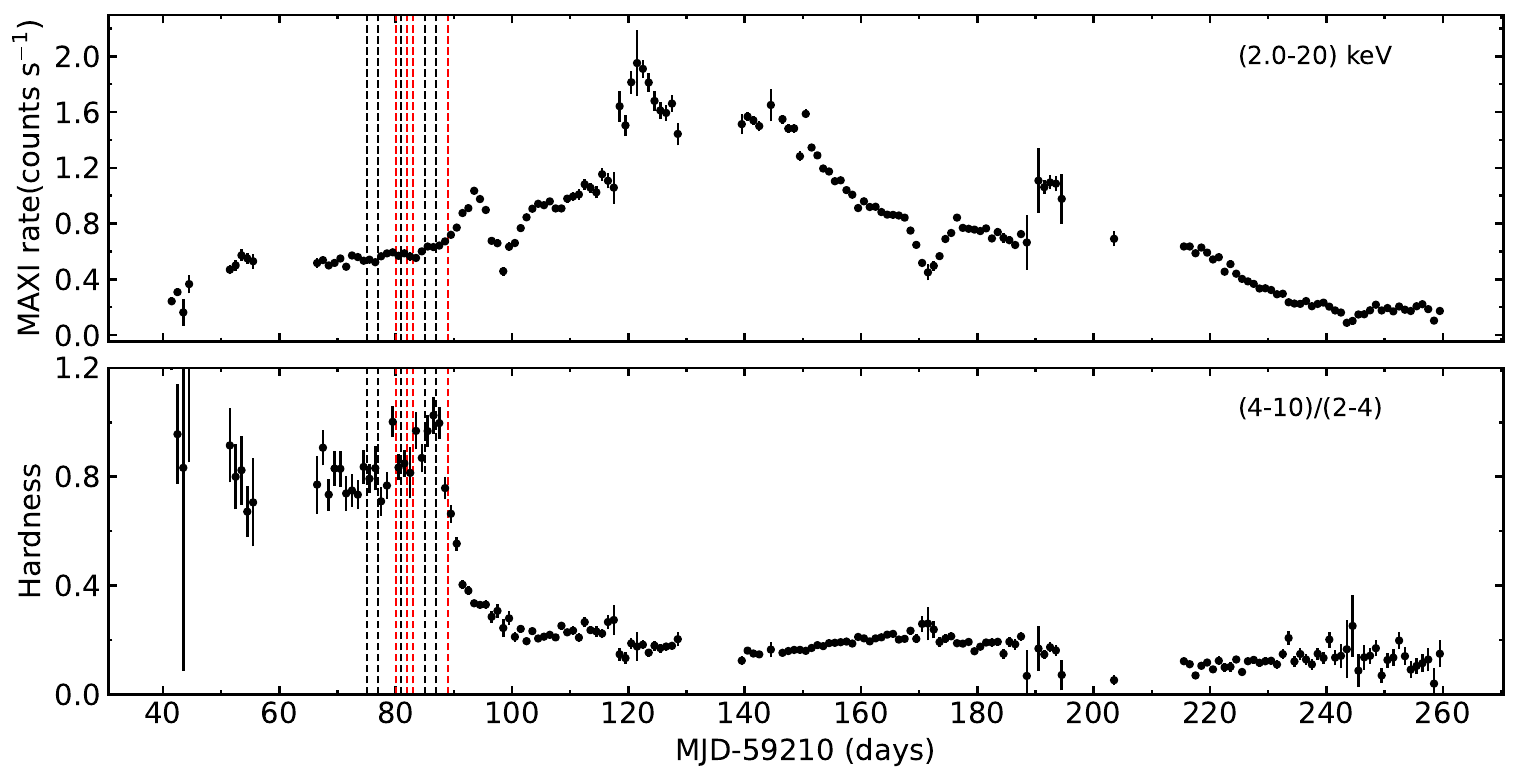} \\
    \caption{\textit{MAXI} GSC light curve and corresponding hardness of \srcg\ starting from 2021 February 5 (MJD = 59250). The hardness is defined as the ratio between the count rates at $4–10$~keV and $2–4$~keV bands. The vertical black lines mark the \hxmt\ observations analyzed in this work and the vertical red lines mark the \nicer\ observations.}
    \label{gx339_maxi}
\end{figure*}


\begin{figure*}
    \centering
    \includegraphics[width=0.9\linewidth]{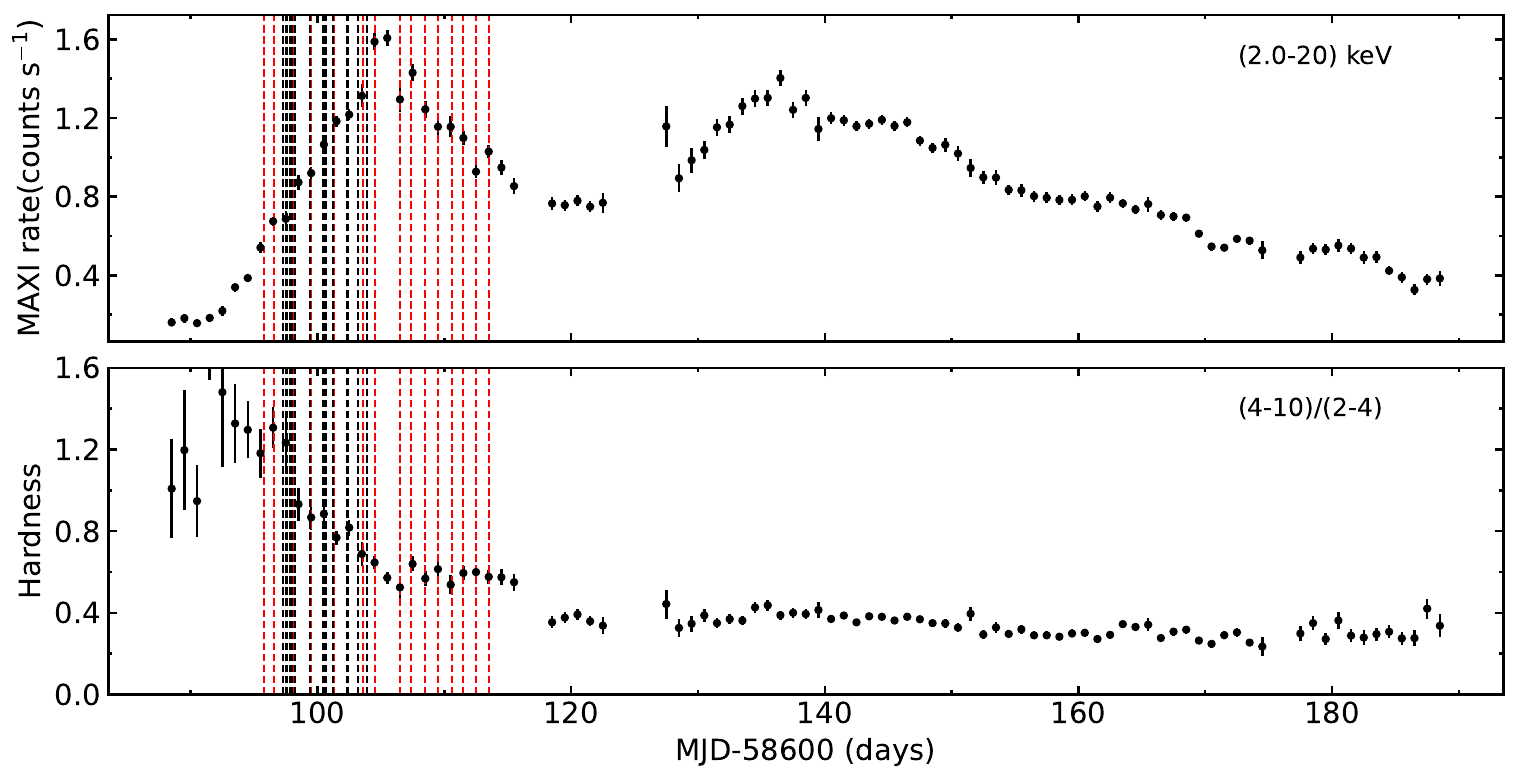} \\
    \caption{\textit{MAXI} GSC light curve and corresponding hardness of \srce\ starting from 2019 June 16 (MJD = 58650). The hardness is defined as the ratio between the count rates at $4–10$~keV and $2–4$~keV bands. The vertical black lines mark the \hxmt\ observations analyzed in this work and the vertical red lines mark the \nicer\ observations.}
    \label{exo1846_maxi}
\end{figure*}


\begin{figure}
    \centering
    \includegraphics[width=0.9\linewidth]{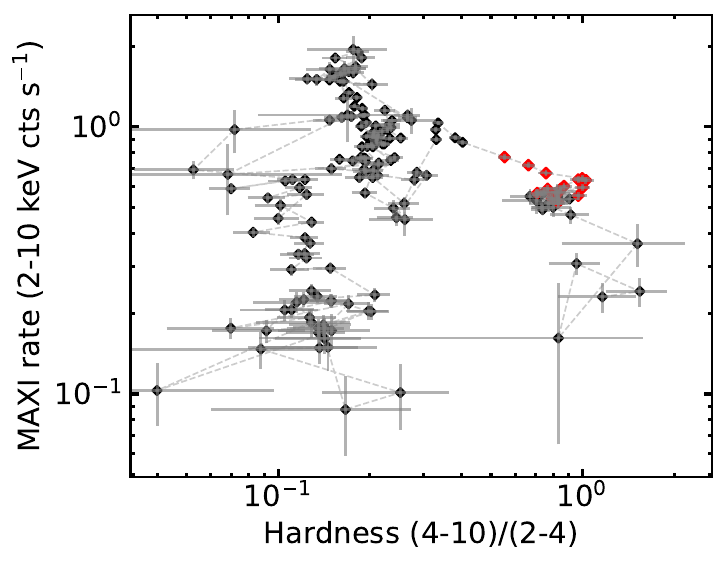} \\
    \caption{Hardness-intensity diagram of \srcg\ with \textit{MAXI} GSC daily average data. The red diamonds represent the \hxmt\ and \nicer\ observations that show type-C QPOs.}
    \label{gx339_hid}
\end{figure}


\begin{figure}
    \centering
    \includegraphics[width=0.9\linewidth]{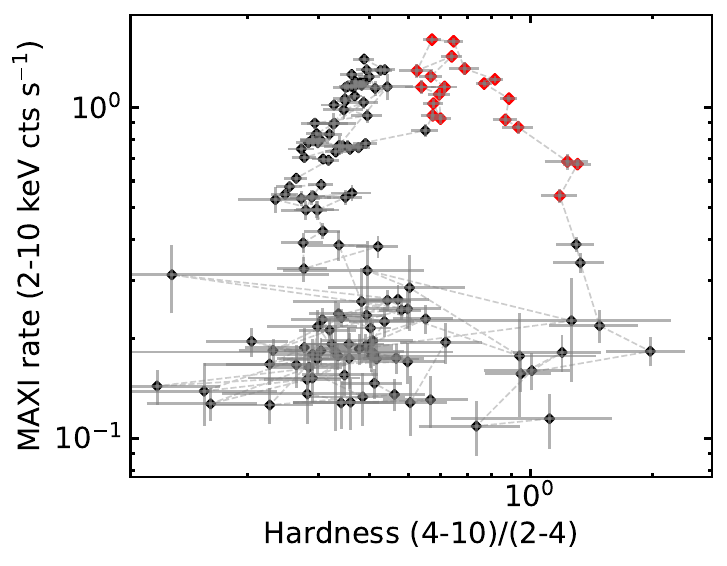} \\
    \caption{Hardness-intensity diagram of \srce\ with \textit{MAXI} GSC daily average data. The red diamonds represent the \hxmt\ and \nicer\ observations that show type-C QPOs.}
    \label{exo1846_hid}
\end{figure}


\section{Observations and data reduction}
\label{obs}

\subsection{Data selection}

Starting from February 2021, \srcg\ went into a new outburst that lasted for a few months. Fig.~\ref{gx339_maxi} shows the long-term light curve in the $2–20$~keV band and the corresponding hardness observed with \textit{MAXI} GSC. The hardness is defined as the ratio between the count rates at $4–10$~keV and $2–4$~keV. Both \hxmt\ and \nicer\ extensively observed the 2021 outburst of the source. We went through all available \hxmt\ and \nicer\ data and picked out those observations that show type–C QPO signatures. Type-C QPOs are characterized by the narrow peak and its harmonic in the PDS, together with a strong flat-top broadband noise. The selected observations analyzed in this work are marked in the light curve of \srcg\ in Fig.~\ref{gx339_maxi}. Information about these observations is listed in Tab.~\ref{gx339_obs}.

The 2019 outburst of \srce\ was first detected by \textit{MAXI}/GSC on 2019 July 23 \citep{Negoro2019}, and it lasted about 3 months. The long-term \textit{MAXI} light curve and the corresponding hardness are shown in Fig.~\ref{exo1846_maxi}. Same as \srcg, both \hxmt\ and \nicer\ conducted high–cadence pointing observations of \srce. Type–C QPOs appear during the transition from hard to soft state \citep{Liu2021}. We selected observations showing type–C QPO signatures. The selected observations are marked in the light curve in Fig.~\ref{exo1846_maxi} and listed in Tab.~\ref{exo1846_obs}.

We show the hardness intensity diagram (HID) of \srcg\ and \srce\ in Fig.~\ref{gx339_hid} and Fig.~\ref{exo1846_hid}, respectively. The red diamonds represent the \hxmt\ and \nicer\ observations that show type-C QPO features.

\subsection{Data reduction}

\hxmt\ covers the broadband energy range of 1–250~keV with low-energy, medium-energy, and high-energy detectors  (\citealt{Zhang2020}; \citealt{Chen2020}; \citealt{Cao2020}; \citealt{Liu2020}). The light curves and spectra are extracted using the \hxmt\ data analysis software (HXMTDAS) version 2.05 and CALDB version 2.06, following the official user guide. The background is estimated by the standalone scripts \texttt{hebkgmap}, \texttt{mebkgmap}, and \texttt{lebkgmap} (\citealt{Guo2020}; \citealt{Liao2020}; \citealt{Liao2020b}). The data are screened following the recommended criteria, i.e., an elevation angle $\textgreater10^{\circ}$, a geomagnetic cutoff rigidity $\textgreater10$~GeV, a pointing offset angle $\textless0.1$, and at least 300~s away from the South Atlantic Anomaly (SAA).

The \nicer\ data are processed with the \nicer\ data analysis software (NICERDAS) version 2021-04-01\_V008 and CALDB version 20210707. We apply standard filtering criteria: the pointing offset is less than $54^{\prime \prime}$, and the pointing direction is more than $40^{\circ}$ away from the bright Earth limb, more than $30^{\circ}$ away from the dark Earth limb, and outside the South Atlantic Anomaly. Additionally, we exclude data from detectors \# 14 and \# 34, which are affected by episodes of increased electronic noise. We select events not flagged as “overshoot” or “undershoot” resets (EVENT\_FLAGS = bxxxx00) or forced triggers (EVENT\_FLAGS=bx1x000). The standard \nicer\ reduction routine \texttt{nicerl2} is used to process the data. The cleaned events are barycenter-corrected using the FTOOL \texttt{barycorr}. We extract the energy spectra of the background in each observation using the \texttt{nibackgen3C50} tool \citep{Remillard2022}. The Redistribution Matrix File (RMF) and Ancillary Response File (ARF) are created by using the tasks \texttt{nicerrmf} and \texttt{nicerarf}, respectively.


\begin{figure*}[]
    \begin{center}
	\includegraphics[scale=1.25]{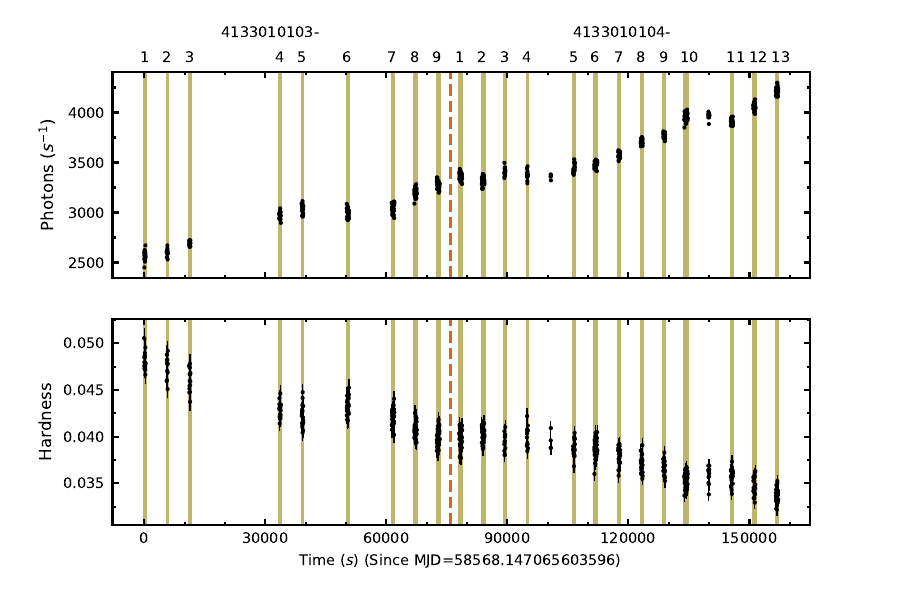}\\
    \end{center}
    \vspace{-0.8cm}
    \caption{Light curves of \srcg\ by \nicer\ in the 0.5–10~keV band (top panel) and corresponding hardness (bottom panel) (ObsID: 4133010103, 4133010104). The hardness is defined as the ratio between the count rates in the $4–10$~keV and $0.5–4$~keV bands. The source undergoes a process of flux increase and hardness decrease during this period. The intervals with exposure $> 200$~s are marked with yellow shadow. The selected sub-observations are numbered 4133010103-1 through 4133010103-9, and 4133010104-1 through 4133010104-13. 
    \label{gx339_nicer_lc}}
\end{figure*}



\section{Data analysis}
\label{analysis}

\subsection{Timing analysis \label{Timing}}

We extract \hxmt\ LE and \nicer\ XTI light curves with a time resolution of 1~ms from the full energy band (1-10~keV in \hxmt; 0.5-10~keV in \nicer) for each \hxmt\ observation and \nicer\ observation. In order to calculate hardness ratios, we also produce LE light curves from the 1-5~keV and 5–10~keV bands, and XTI light curves from the 0.5-4~keV and 4-10~keV bands. 

We carefully check the extracted light curves from all observations of \srcg\ and find there are two \nicer\ observations (ObsID: 4133010103, 4133010104) that show relatively strong variability in count rate and hardness. Fig.~\ref{gx339_nicer_lc} shows the light curves of these two \nicer\ observations. The gaps in the light curves are due to the low Earth orbit of the telescope or the SAA. We can clearly see that the source went through a period of luminosity increase and hardness decrease. Comparing with the location of these two observations in Fig.~\ref{gx339_maxi} (the last red dotted line; in fact, since the two lines are quite close, they look like one line), we conclude that these two observations are during the hard-to-soft transition with the hardness continuously decreasing. Then we divide the observations according to the light curve, counting each snapshot as a sub-observation. After checking all sub-observation data, we select those with exposures greater than $200$~s. The selected sub-observations are numbered 4133010103-1 through 4133010103-9, and 4133010104-1 through 4133010104-13, as shown in Fig.~\ref{gx339_nicer_lc}. The other light curves do not show strong variability in the count rate i.e., no distinctive evidence of flares, dips, or state transitions, ensuring a safe basis for timing and spectral analysis to characterize the source properties.


\begin{figure*}[t]
    \begin{center}
       \includegraphics[width=0.9\linewidth]{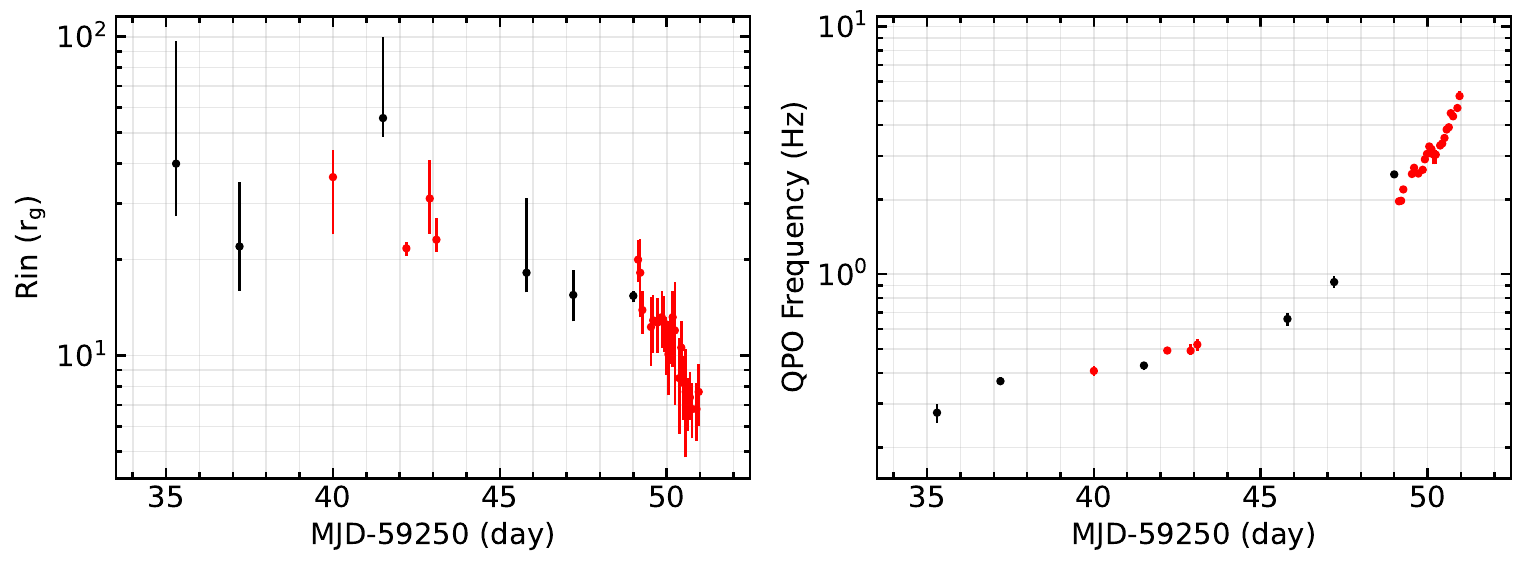}
		\\
    \end{center}
    \vspace{-0.2cm}
    \caption{Evolution of disk inner radius $R_{\rm in}$ (left panel) and QPO frequency ($f_{\rm QPO}$) along with MJD in \srcg. The black points indicate \hxmt\ data, and red dots indicate \nicer\ data. 
    \label{gx339_rin_freq_mjd}}
\end{figure*}


For \srce, the count rate of the source remain fairly stable during each \hxmt\ and \nicer\ interval, and the hardness does not change dramatically. Therefore, we carry out timing and spectral analysis in the unit of one observation in this work.

To measure the QPO frequency ($f_{\rm QPO}$) of \srcg\ and \srce, we employ the \texttt{Python} package \texttt{Stingray} \citep{Huppenkothen2019} to create PDS for each observation. The light curve of both \hxmt\ and \nicer\ are split into 64~s segments, and then PDSs from all 64s segments are averaged. The PDS is normalized according to the "rms" method \citep{Belloni1990}, and logarithmically rebinned so that each bin size is 1.02 times larger than the previous bin. Note that we focus on the \hxmt\ LE 1-10~keV light curve and the \nicer\ XTI 0.5-10~keV light curve to extract PDS and search for QPO signal. The time resolution of both \hxmt\ LE and \nicer\ XTI lightcurve is 1~ms, corresponding to the Nyquist frequency of 500 Hz. The 8–30~keV \hxmt\ ME light curves have been analyzed in the same way and return consistent measurements of the QPO frequencies. So we report only the results from LE data in this work.

We use XSPEC v12.12.1 \citep{Arnaud1996} to analyze the PDS. The typical PDS of XRBs manifests broad components and one or two narrow peaks at different frequencies, corresponding to broad band noise, the possible QPO fundamental and harmonics, respectively. We need at least one narrow Lorentzian for the QPO to fit the Poisson-extracted PDS \citep{Belloni2002}. More narrow Lorentzians are sometimes included to model harmonic peaks. All QPOs we detect have a quality factor (Q) greater than 4 and detection significance greater than $3\sigma$, i.e., the ratio of the Lorentzian norm divided by its $1\sigma$ negative error is larger than 3. Fig.~\ref{hxmt_gx339} and Fig.~\ref{nicer_gx339} show a typical PDS and the fit results with several Lorentzian models for \srcg. Fig.~\ref{hxmt_exo1846} and Fig.~\ref{nicer_exo1846} show the counterpart for \srce. The QPO frequencies for each observation are listed in Tab.~\ref{result_gx339} for \srcg, and Tab.~\ref{result_exo1846} for \srce.



\begin{figure*}[]
    \begin{center}
       \includegraphics[width=0.9\linewidth]{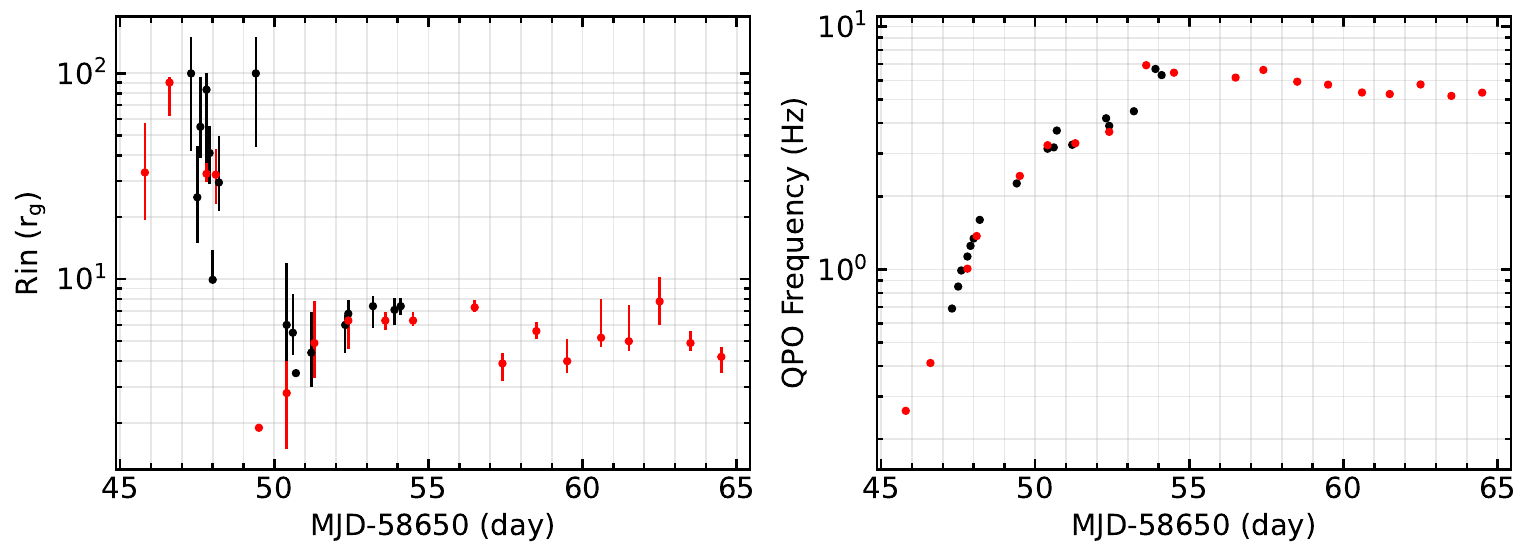}
	\\
    \end{center}
    \vspace{-0.2cm}
    \caption{Evolution of disk inner radius $R_{\rm in}$ (left panel) and QPO frequency ($f_{\rm QPO}$) along with MJD in \srce. The black points and red points represent the results of \hxmt\ data and \nicer\ data, respectively.
    \label{exo1846_rin_freq_mjd}}
\end{figure*}



\subsection{Spectral analysis \label{Spectral}}

For spectral analysis of the \hxmt\ data, we consider the LE data in the 2–10~keV band and ME data in the 8–20~keV band. ME data above 20~keV and HE data are ignored because of the very high background. Note that we ignore the data below 2~keV of the LE instrument in spectral analysis (instead of the 1~keV for timing analysis) because of calibration uncertainties in the low energy band. For \nicer\ data, we consider the 1-10~keV band in this section, ignoring the data below 1~keV because of calibration issues.

The \hxmt\ and \nicer\ spectra are fitted with the XSPEC (v12.12.1) package. The $\chi^2$ statistics is employed and all parameter uncertainties are estimated at 90\% confidence level, corresponding to $\Delta\chi^2=2.71$. All spectra are grouped to ensure a minimum counts of 20 per bin. A systematic error of 1\% is added to the \nicer\ spectra.

The \hxmt\ and \nicer\ spectra of \srcg\ are fitted with the model combination \texttt{Tbabs $\times$ (simpl $\times$ kerrd + relxill)}. \texttt{Tbabs} is included to account for absorption by the interstellar medium. The recommended photoelectric cross sections of \citet{Verner1996} and element abundances of \citet{Wilms2000} are used for the \texttt{Tbabs} component. We set its column density ($n_{\rm H}$) to be a free parameter for \nicer\ spectra. While with \hxmt\ spectra, we can not well constrain $n_{\rm H}$, so we fix it at best-fit value, $0.55 \times 10^{22}$~cm$^{-2}$, which is consistent with the result of the \nicer\ data and the value in literature (e.g., \citealt{Wang2020}; \citealt{Liu2022}). \texttt{kerrd} accounts for the thermal emission from the geometrically thin and optically thick accretion disk \citep{Ebisawa2003}, in which the black hole distance, mass, and inclination angle of the accretion disk are set to 8.4~kpc, 9.0~M$_{\bigodot}$, and 30$^\circ$ \citep{Parker2016}, respectively. The spectral hardening factor of \texttt{kerrd} is set to 1.7 \citep{Shimura1995}. \texttt{simpl} \citep{Steiner2009} is used to take into account for the Comptonization of disk photons by the corona. The source has been found to have strong reflection features \citep{Liu2022}, therefore we use the full reflection model \texttt{relxill} \citep{Garcia2014} to fit them. The spin parameter ($a_*$) is fixed at 0.95 \citep{Parker2016}. The index of the emissivity profile is fixed at 3 because it cannot be constrained by the fit. The iron abundance ($A_{\rm Fe}$) and reflection fraction are fixed at 1 and -1, respectively. The best-fit values and uncertainties are shown in Tab.~\ref{result_gx339}. Fig.~\ref{hxmt_gx339} and Fig.~\ref{nicer_gx339} show typical spectra and fit results of \hxmt\ data and \nicer\ data, respectively.

In the case of \srce, for \hxmt\ spectra, the best-fit model is \texttt{Tbabs $\times$ (simpl $\times$ kerrd + relxill)}. The black hole distance, mass, and inclination angle of the accretion disk are set to 4.5~kpc \citep{Williams2022}, 10.0~M$_{\bigodot}$ (\citealt{Williams2022}, \citealt{Draghis2020}) and 73$^\circ$ \citep{Draghis2020}, respectively. The spin $a_*$ is fixed at 0.998 \citep{Draghis2020}. We use a simple power-law to model the emissivity profile ($q_{\rm in}=q_{\rm out}$ free). The other parameters are set exactly as in the case of \srcg. For \nicer\ spectra, we notice that there are still some large residuals in the soft X-ray band with the same model, including a Gaussian-like emission near 1.1~keV and edge-like shapes near~1.8 keV. These energies correspond to features in the effective area of \nicer\ versus energy (e.g., \citealt{Wang2020}), where 1.1 and 1.8~keV are attributed to sodium and silicon, respectively. Therefore, we adopt the following model for the \nicer\ spectra: \texttt{Tbabs $\times$ (simpl $\times$ kerrd + relxill + gaussian) $\times$ edge}. This calibration issue arises in \srce\ because the source has high interstellar absorption, which makes the photon count rate in the lower energy band relatively low, making the calibration issue prominent. Typical spectra and fit results of \hxmt\ and \nicer\ are shown in Fig.~\ref{hxmt_exo1846} and Fig.~\ref{nicer_exo1846}. In Tab.~\ref{result_exo1846}, we summarize the best-fit values and errors of \srce.



\begin{figure*}[]
    \begin{center}
       \includegraphics[width=0.9\linewidth]{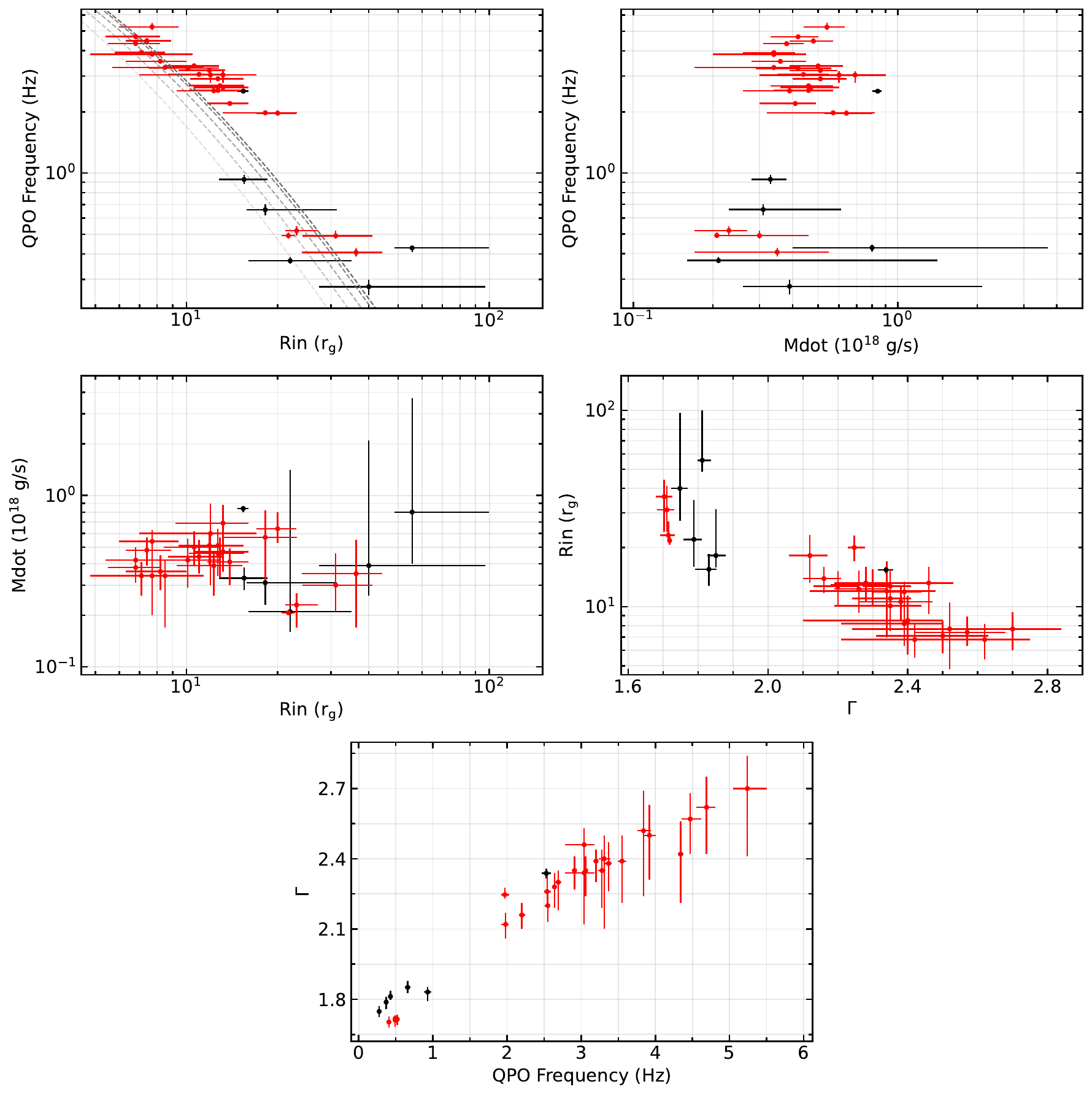}
	\\
    \end{center}
    \vspace{-0.2cm}
    \caption{Correlation between the parameters involved in the temporal and spectral analysis in the case of \srcg. Correlation of the QPO frequency vs. inner disk radius and the QPO frequency vs. accretion rate are shown in the upper left and upper right panels. The two central panels illustrate the accretion rate vs. inner disk radius and inner disk radius vs. photon index. The photon index vs. QPO frequency is depicted in the bottom panel. The black and red crosses denote the results of \hxmt\ data and \nicer\ data, respectively. In the left top panel, the dashed gray lines represent the correlation of the frequency and inner radius predicted by Lense–Thirring precession model. The lines from left to right depict $a_* = 0.3$, 0.5, 0.7, 0.9 and 0.998, respectively. 
    \label{gx339_qpo_spectra_correlation}}
\end{figure*}



\begin{figure*}[]
    \begin{center}
	\includegraphics[width=0.9\linewidth]{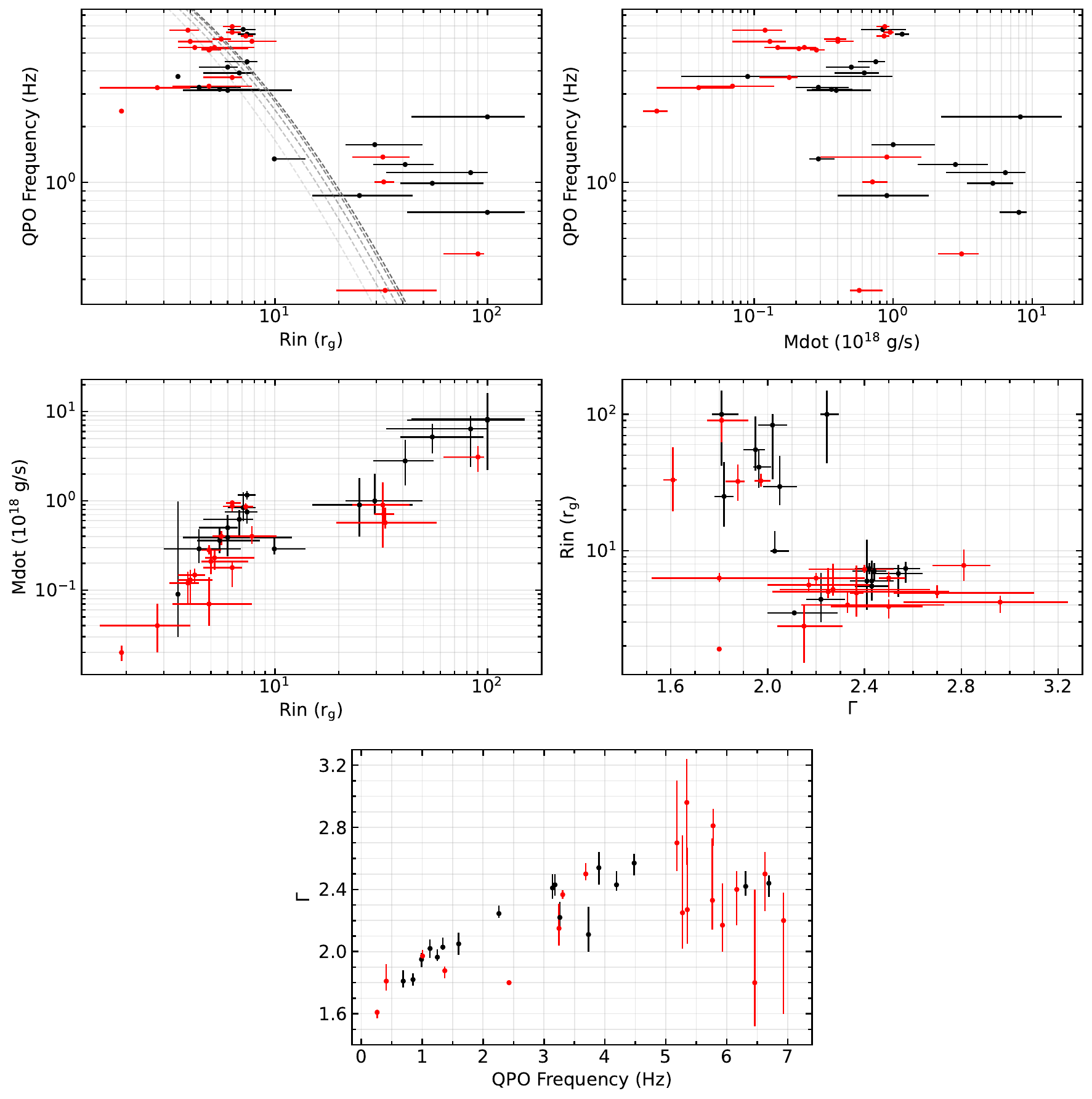}
        \\
    \end{center}
    \vspace{-0.2cm}
    \caption{Correlation between the parameters involved in temporal and spectral analysis in the case of \srce. It is organized as in Fig.~\ref{gx339_qpo_spectra_correlation}.
    \label{exo1846_qpo_spectra_correlation}}
\end{figure*}


\begin{figure}[]
    \begin{center}
       \includegraphics[width=0.9\linewidth]{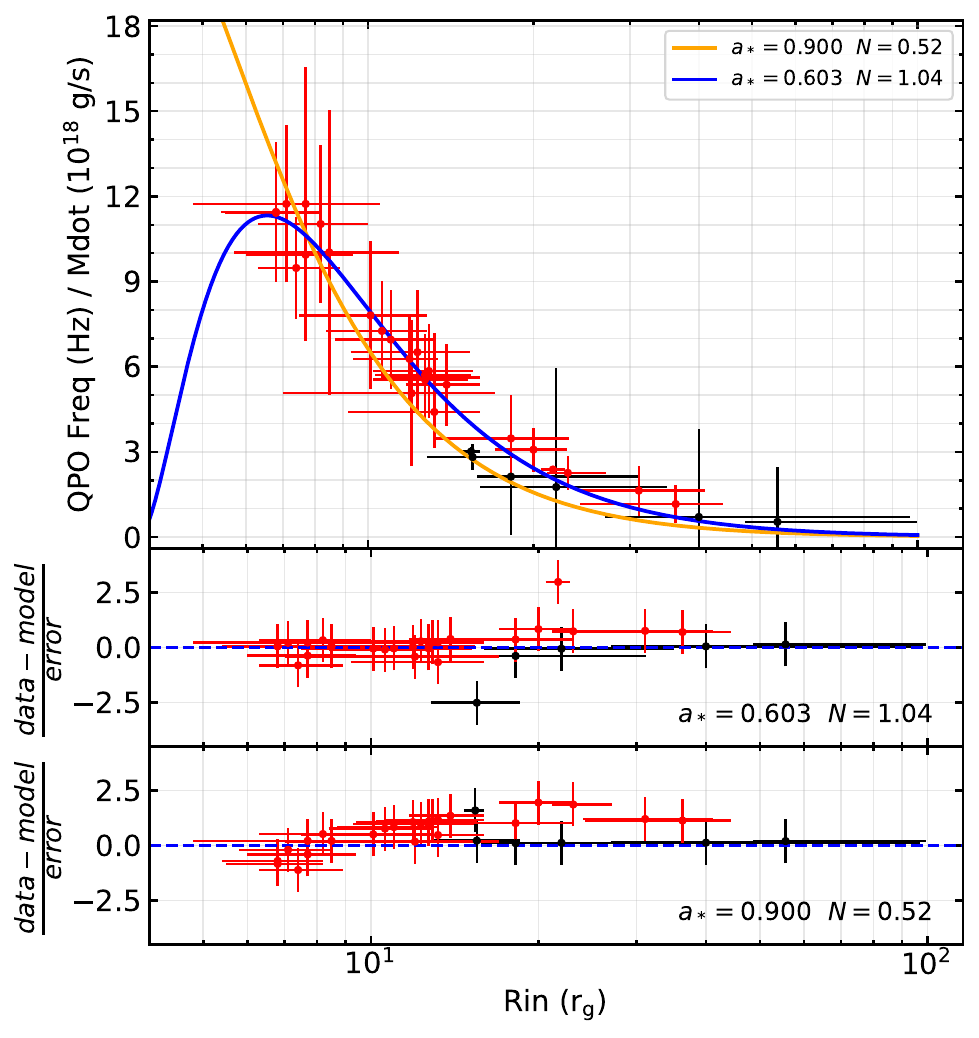}
	\\
    \end{center}
    \vspace{-0.2cm}
    \caption{Top: variation of QPO frequency divided by the accretion rate vs. inner disk radius in the case of \srcg. The results of \hxmt\ data and \nicer\ data are denoted with black and red crosses, respectively. The orange curve represents the best fit ($a_* = 0.603$, $N = 1.02$), and the blue curve corresponds to $a_* = 0.900$, $N = 0.52$. Middle: residuals of the model ($a_* = 0.603$, $N = 1.02$) to data. Bottom: residuals of the model ($a_* = 0.900$, $N = 0.52$) to data. In the middle and bottom panels, the y-axes are scaled for clarity and there is, respectively, 1 and 2 points that are outside the plots.
    \label{gx339_dyn}}
\end{figure}




\begin{figure}[]
   \begin{center}
	\includegraphics[width=0.95\linewidth]{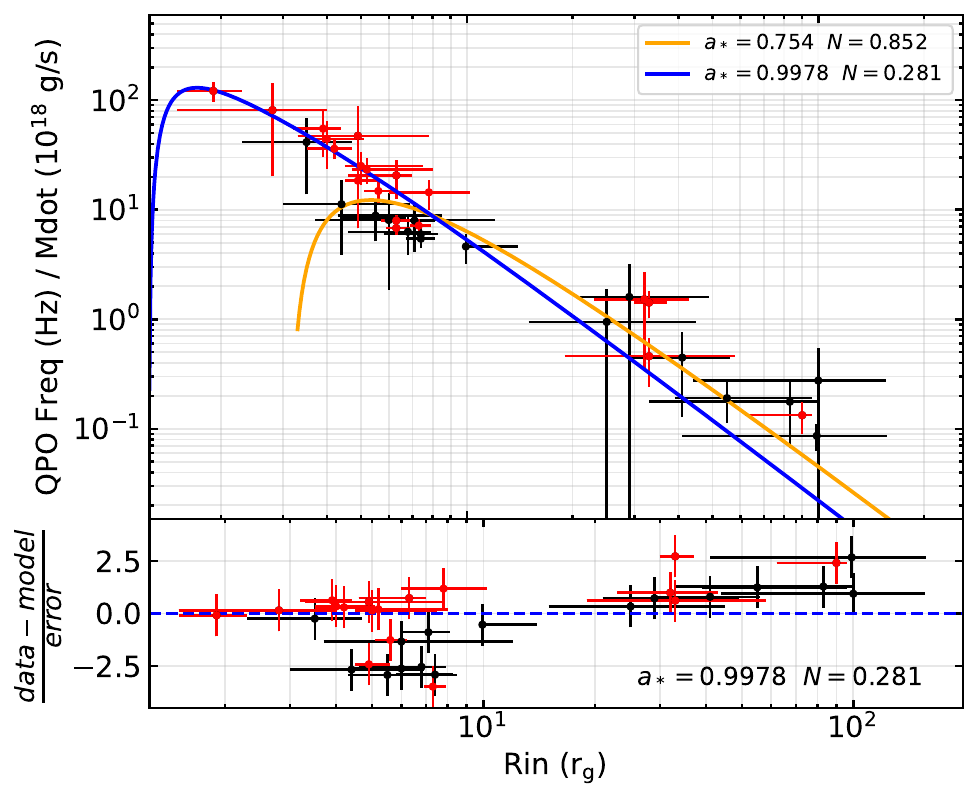}
	\\
   \end{center}
   \vspace{-0.2cm}
   \caption{Top: variation of QPO frequency divided by the accretion rate vs. inner disk radius in the case of \srce. The black and red crosses denote the results of \hxmt\ data and \nicer\ data, respectively. The blue curve shows the best fit. The orange curve represents the best fit ($a_* = 0.754$, $N = 0.853$) when we only include the data with $R_{\rm in} > 5$~$R_{\rm g}$. It proves that the data with small $R_{\rm in}$ are important to fit the spin parameter. Bottom: residuals of the best-fit model ($a_* = 0.9978$, $N = 0.281$) to data. The y-axis is scaled for clarity and there are 3 points that are outside the plots.
   \label{exo1846_dyn}}
\end{figure}



\section{Results and discussion}
\label{discuss}

Fig.~\ref{gx339_rin_freq_mjd} and Fig.~\ref{exo1846_rin_freq_mjd} show the evolution of inner radius ($R_{in}$) and $f_{\rm QPO}$ with time for \srcg\ and \srce, respectively. Generally speaking, we clearly see that the value of $f_{\rm QPO}$ monotonically increases with time. The behaviour is consistent with that reported in \citet{Motta2011} and \citet{Liuhx2021}. It has also been observed in other XRBs, for example, XTE~J1859+226 \citep{Casella2004}. In addition, a notable feature for both sources is the decrease of $R_{in}$. For \srcg, the inner disk moves toward the ISCO (Innermost Stable Circular Orbit), from $> 50 R_{\rm g}$ to $\sim 7 R_{\rm g}$ ($R_{\rm g}$, gravitional radius), which coincides with the result of previous study (e.g. \citealt{Wang-ji2018}; \citealt{Wang2020}). Although there are some variable features, \srce\ shows a similar trend. Left panels of Fig.~\ref{gx339_rin_freq_mjd} and Fig.~\ref{exo1846_rin_freq_mjd} show a break of the relation for \srcg\ and \srce, respectively. Same for the QPO frequency on the right panels. This break of the relation is most likely related to the transition from hard state to hard intermediate state, as suggested in \citet{Wang2022}.

Correlation between the parameters involved in temporal and spectral analysis are shown in Fig.~\ref{gx339_qpo_spectra_correlation} and Fig.~\ref{exo1846_qpo_spectra_correlation}. An interesting result is the relationship between the photon index ($\Gamma$) and $f_{\rm QPO}$. The results from both sources exhibit the same tendency, as shown in the bottom panels of Fig.~\ref{gx339_qpo_spectra_correlation} and Fig.~\ref{exo1846_qpo_spectra_correlation}. There is a significant positive correlation between $\Gamma$ and $f_{\rm QPO}$ initially, which flattens or starts reversing at the highest values of the $f_{\rm QPO}$. The turnoff in the correlation is not apparent in \srcg, while it is evident in \srce\ (around $\Gamma \sim 2.7$). A similar correlation have been reported in a number of other LMXBs (e.g., \citealt{Vignarca2003}; \citealt{Titarchuk2004}; \citealt{Titarchuk2009}; \citealt{Furst2016}). \citet{Titarchuk2004} introduced the transition layer (TL) model to explain the observed correlations. The TL model depicts how the QPOs related to the corona properties (e.g., the size, optical depth, temperature and spectral index), and predicts the correlation between photon index and QPO frequency. The results we get are in significant agreement with the predictions of model, except for the observations of \srce\ with $f_{\rm QPO} > 5.18$~Hz, where a negative correlation between $f_{\rm QPO}$ and $\Gamma$ appears. A universal explanation of this correlation between $\Gamma$ and $f_{\rm QPO}$ is still missing.

The upper left panel of Fig.~\ref{gx339_qpo_spectra_correlation} shows a broad anti-correlation between $f_{\rm QPO}$ and $R_{in}$ in \srcg. This anti-correlation is not particularly significant in \srce, and we can only observe a general tendency of larger $R_{in}$ corresponding to a smaller frequency, as shown in the upper left panel of Fig.~\ref{exo1846_qpo_spectra_correlation}. The same correlation between the QPO frequency and the disk inner radius was reported in other sources (e.g., GRS~1915+105; \citealt{Rodriguez2002}). The hot flow Lense–Thirring precession model would predict anti-correlation between $f_{\rm QPO}$ and $R_{in}$ (\citealt{Ingram2009}; \citealt{Ingram2010}). To check the possibility of modeling the results with the relativistic precession model, we plot the precession frequency calculated with the equation 2 in \citet{Ingram2009}, assuming $a_* = 0.3$, 0.5, 0.7, 0.9 and 0.998, respectively. In these calculations, we choose $\zeta = 0$ and inner radius of hot flow $r_i = 3.0(h/r)^{-4/5}a_*^{2/5}$ (with $h/r = 0.2$), following \citet{Ingram2009}. We see the model cannot explain the results we obtained, neither for \srcg\ nor for \srce, as shown in the plot.

The variation of the QPO frequency with the accretion rate ($\dot{M}$) are shown in the upper right panels of Fig.~\ref{gx339_qpo_spectra_correlation} and Fig.~\ref{exo1846_qpo_spectra_correlation}. \citet{Liu2021} reported a strong correlation between the QPO frequency and mass accretion rate in GRS~1915+105 with \hxmt\ data. We do not find any significant correlation between them in \srcg, while there is a weak anti-correlation in \srce. In fact, a positive correlation between $f_{\rm QPO}$ and $\dot{M}$ is proposed in the TL model by \citet{Titarchuk2004}, and in the disk-corona natural frequency model by \citet{Mastichiadis2022}. Fig. 3 of \citet{Titarchuk2004} depicts the positive correlation between $f_{\rm QPO}$ and the $\gamma$-parameter (which is proportional to mass accretion rate), which is opposite to what we find. \citet{Mastichiadis2022} argued that type-C QPOs could arise from the interaction of the hot corona with the cold accretion disk, and predict a formula $f_{\rm 0} \propto \dot{M}^{1/2}$, for a certain radius of the corona and ratio of hard-to-soft radiation (Fig. 5 of \citet{Mastichiadis2022}). This model possibly reproduce the relation between the $f_{\rm QPO}$ and $\dot{M}$ of our cases, allowing for continuous variations of parameters, such as the size of corona and corona-disk radiative feedback. Our spectral fits do not provide the information of these parameters. So we do not explore further such a possibility.

\citet{Misra2020} identified QPOs as the dynamic frequency of a truncated relativistic accretion disk in the case of GRS~1915+105. The dynamic frequency is defined as the ratio of the sound propagation velocity of the inner disk to the truncation radius, i.e., the inverse of the sound crossing time \citep{Misra2020}. Based on the assumption that the accretion disk is a standard relativistic accretion disk \citep{Novikov1973}, the dynamic frequency is a function of $R_{\rm in}$, $a_{*}$, $\dot{M}$ and a normalization factor ($N$). \citet{Liu2021} extended the results in \citet{Misra2020} to a larger range of accretion rates with \hxmt\ data of GRS~1915+105, and confirmed the high spin nature of the source. Following the work of \citet{Misra2020} and \citet{Liu2021}, we illustrate the relation between $f_{\rm QPO}$ divided by $\dot{M}$ and $R_{\rm in}$ in Fig.~\ref{gx339_dyn} and Fig.~\ref{exo1846_dyn}. Both sources show negative correlation between $f_{\rm QPO}/\dot{M}$ and $R_{\rm in}$, Moreover, the correlation follows the prediction of the dynamic frequency model. 

We fit the relation between $f_{\rm QPO}/\dot{M}$ and $R_{\rm in}$ using Equation (3) in \citet{Misra2020}. We calculate $\chi^2$ as $$\chi^2=\sum_{n=1}^{N}\frac{(y_i-f(x_i))^2}{\delta_i^2}$$ where $\delta_i^2=(\partial f/\partial x)^2(\delta x_i)^2+(\delta y_i)^2$, $\delta x_i$ and $\delta y_i$ are the error of $x_i$ and $y_i$, respectively. The fit returns $a_* = 0.9978 \pm 0.0009$ and $N = 0.281 \pm 0.025$ for \srce, indicating a rapidly spinning black hole. This result is consistent with what has been reported by analyzing the blurred reflection spectra (e.g., \citealt{Draghis2020}; \citealt{Abdikamalov2021}). The best-fit curve is shown in Fig.~\ref{exo1846_dyn}. Note that in the plots for residuals, we only consider the contribution of the error of y-axis.

In the case of \srcg, the fit returns $a_* = 0.603 \pm 0.026$ and $N = 1.02 \pm 0.05$. Such a low spin result is somewhat different from the result obtained by analyzing the blurred reflection spectra or thermal spectra (e.g. \citealt{Reis2008}; \citealt{Ludlam2015}; \citealt{Garcia2015}; \citealt{Parker2016}; \citealt{Wang2020}). We note that for this source we do not have data below 6~$R_{\rm g}$. The relativistic effects are more evident at lower $R_{\rm in}$ (3 $\sim$ 5~$R_{\rm g}$). Hence, data points at lower $R_{\rm g}$ plays a crucial role in the estimation of spin parameter value. In Fig.~\ref{gx339_dyn}, we simultaneously show two curves, $a_* = 0.603$ and $a_* = 0.900$.  It is worth noting that the most important difference between the two curves is reflected in the region with low $R_{\rm in}$. This also proves our view that a reasonable fitting value cannot be obtained because of the lack of data with relatively small $R_{\rm in}$.

The evolution of QPO frequency is generally consistent with the prediction of the dynamic frequency model. However, the residuals still show a trend: negative values always appear at low $R_{\rm in}$ while positive values appear at large $R_{\rm in}$. We note that the dynamic frequency model is a very simple assumption. It might not be able to capture all of the factors that drive the variability of the LFQPOs. Therefore it is reasonable that we find some differences between different observations.

The middle right panels of Fig.~\ref{gx339_qpo_spectra_correlation} and Fig.~\ref{exo1846_qpo_spectra_correlation} show that the inner disk radius tends to decrease when $\Gamma$ increases. The behaviour is consistent with that expected during a hard-to-soft transition. A noteworthy positive correlation between $\dot{M}$ and $R_{in}$ in \srce\ is described in the middle left panel of Fig.~\ref{exo1846_qpo_spectra_correlation}. A similar relationship was reported in GRS~1915+105 (\citealt{Misra2020}; \citealt{Liu2021}; \citealt{Rawat2022}) and MAXI~J1535-571 \citep{Garg2022}. This correlation does not follow the expected behavior of the truncated disk model \citep{Done2007}. But \citet{Dullemond2005} predicted a positive correlation between $\dot{M}$ and $R_{\rm in}$ (see their Fig. 8), calculating the evaporation of the cool accretion disk on account of the ion-bombardment. An alternative explanation is discussed in \citet{Abramowicz1978}, where the authors suggested a larger inner edge is required when the mass accretion rate increases to dissipate the angular momentum of accretion material. For these models, a detailed study is beyond the scope of this work, and we leave it for future work.


\section{Conclusion}
\label{conclusion}

We investigate the temporal and spectral properties from the latest \hxmt\ and \nicer\ observing campaign of \srcg\ and \srce. Temporal and spectral analyses show that the evolution of the QPO frequency is closely related to mass accretion rate and inner disk radius in both sources, and are generally consistent with the prediction of the dynamic frequency model. We extend the application of the model from GRS~1915+105, a persistent source, to these two transient sources, and confirm the high spin nature of the black hole in \srce.



\section*{Acknowledgements}

This work was supported by the National Natural Science Foundation of China (NSFC), Grant No.~12250610185, 11973019, and 12261131497, the Natural Science Foundation of Shanghai, Grant No.~22ZR1403400, and the Shanghai Municipal Education Commission, Grant No.~2019-01-07-00-07-E00035.


\appendix
\renewcommand\thefigure{\Alph{section}\arabic{figure}}  
\renewcommand\thetable{\Alph{section}\arabic{table}}  

\section{Appendix~A}
\setcounter{figure}{0}  
\setcounter{table}{0}

\begin{figure*}[]
    \begin{center}
	\includegraphics[width=0.4\linewidth]{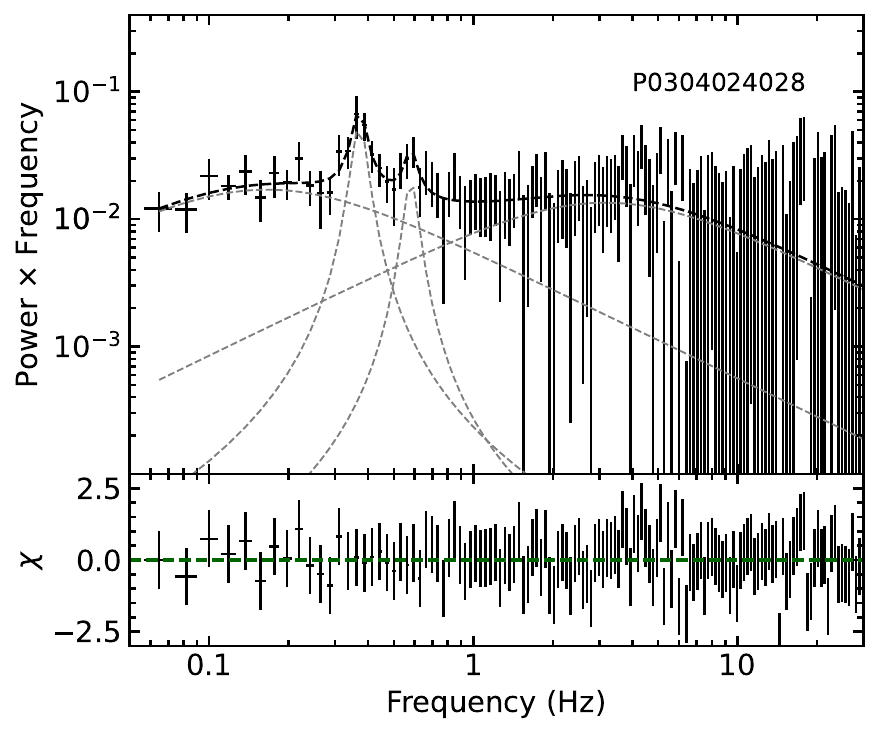}
	\includegraphics[width=0.4\linewidth]{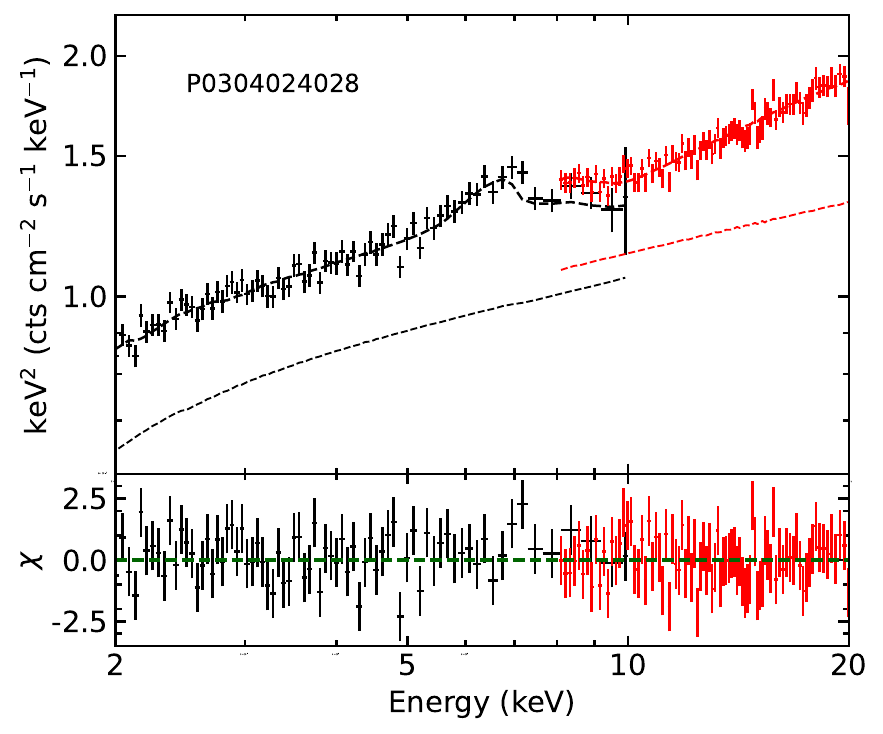} \\ 
    \end{center}
    \vspace{-0.2cm}
    \caption{Left: a typical PDS of \srcg\ from \hxmt\ data (Obs ID: P0304024028). Right: the \hxmt\ spectrum and residuals to the best-fit model for the same observation. Data from the LE and ME detector are denoted in black and red, respectively. 
    \label{hxmt_gx339}}
\end{figure*}

\begin{figure*}[]
    \begin{center}
	\includegraphics[width=0.4\linewidth]{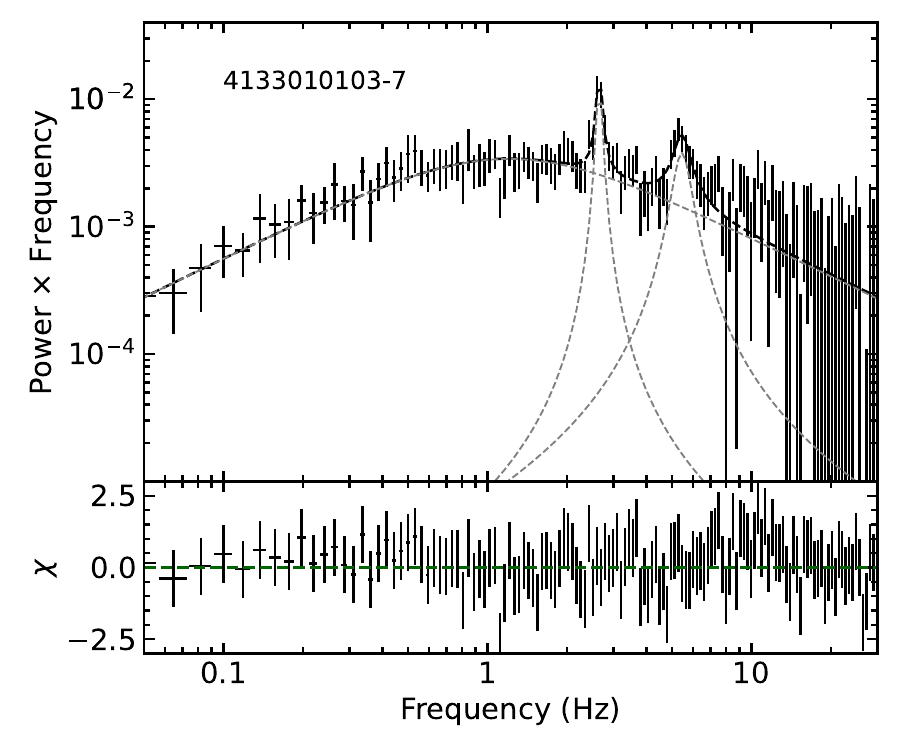}
	\includegraphics[width=0.4\linewidth]{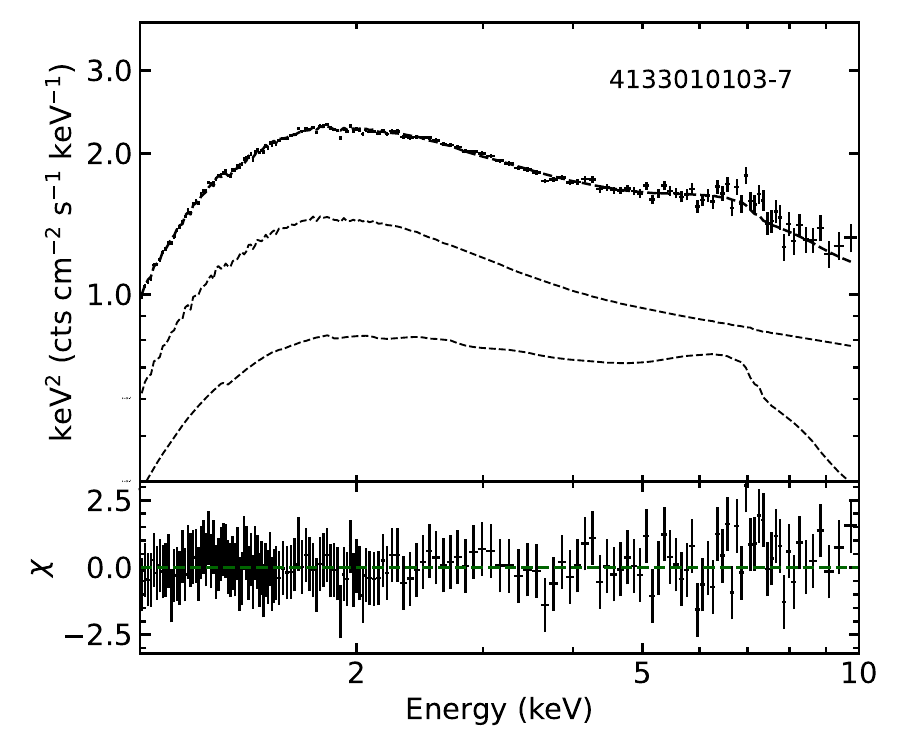} \\ 
    \end{center}
    \vspace{-0.2cm}
    \caption{Left: a typical PDS of \srcg\ from \nicer\ data (Obs ID: 4133010103-7). Right: the \nicer\ spectrum and residuals to the best-fit model for the same observation. 
    \label{nicer_gx339}}
\end{figure*}

\begin{figure*}[]
    \begin{center}
	\includegraphics[width=0.4\linewidth]{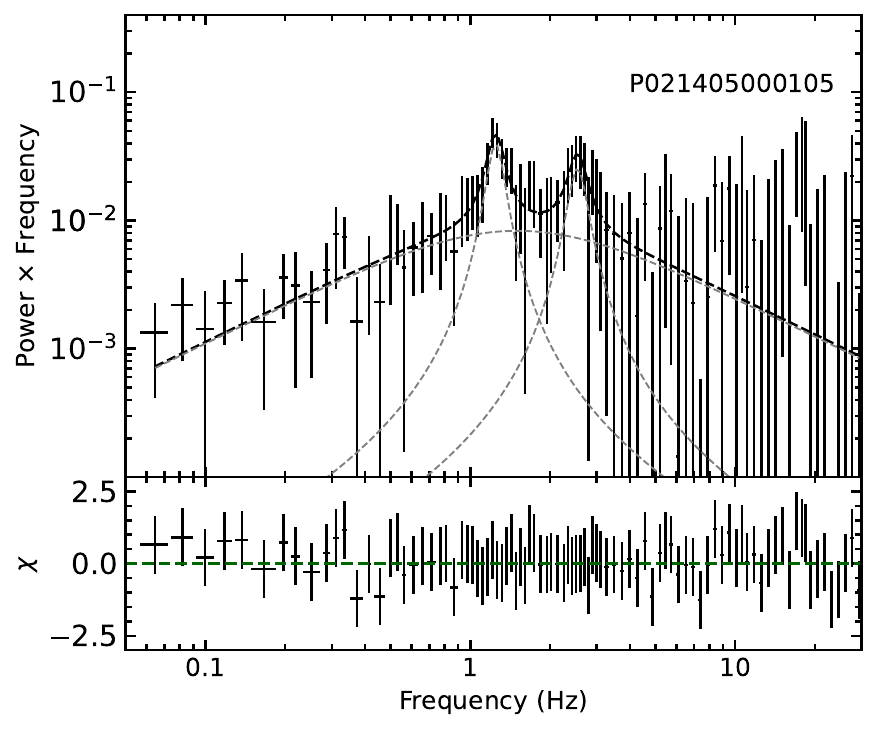}
	\includegraphics[width=0.4\linewidth]{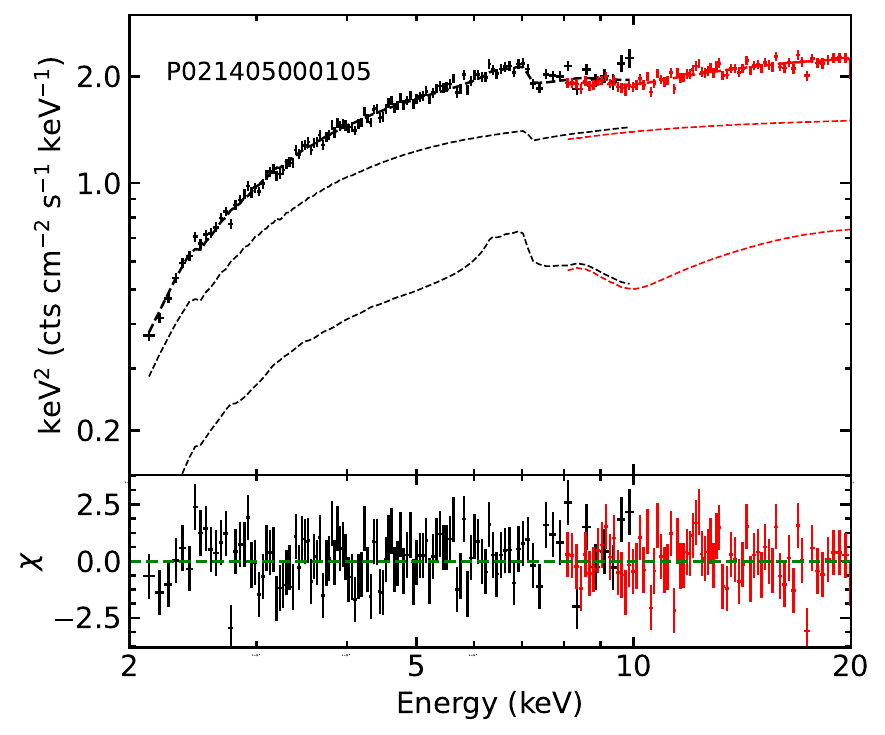} \\ 
    \end{center}
    \vspace{-0.2cm}
    \caption{Left: a typical PDS of \srce\ from \hxmt\ data (Obs ID: P021405000105). Right: the \hxmt\ spectrum and residuals to the best-fit model for the same observation. Same as Fig.~\ref{hxmt_gx339}, data from the LE and ME detector are denoted in black and red, respectively. 
    \label{hxmt_exo1846}}
\end{figure*}

\begin{figure*}[]
    \begin{center}
	\includegraphics[width=0.4\linewidth]{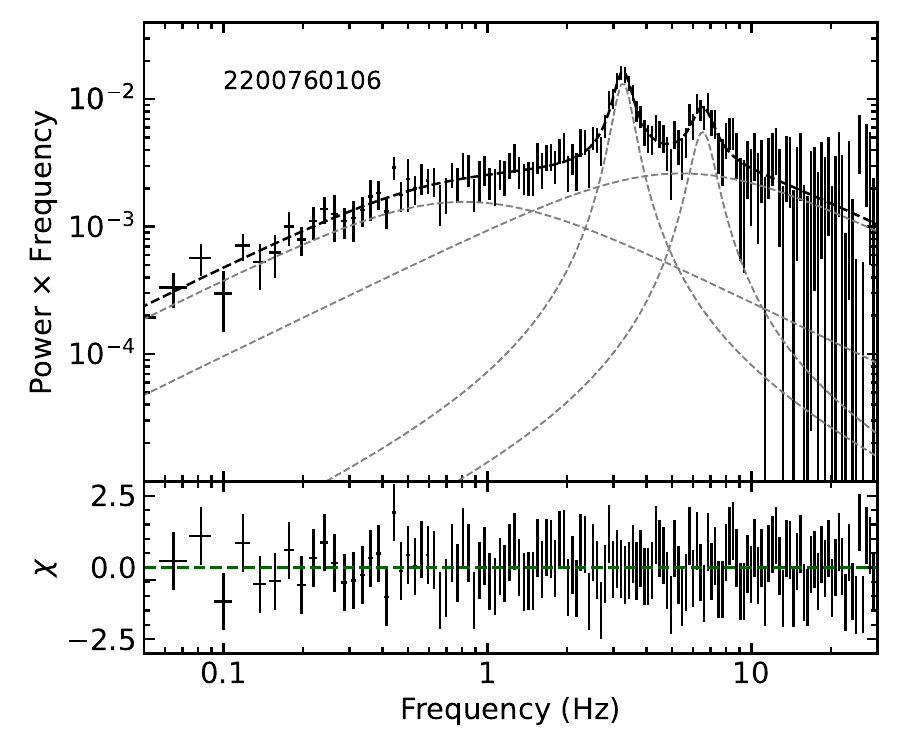}
	\includegraphics[width=0.4\linewidth]{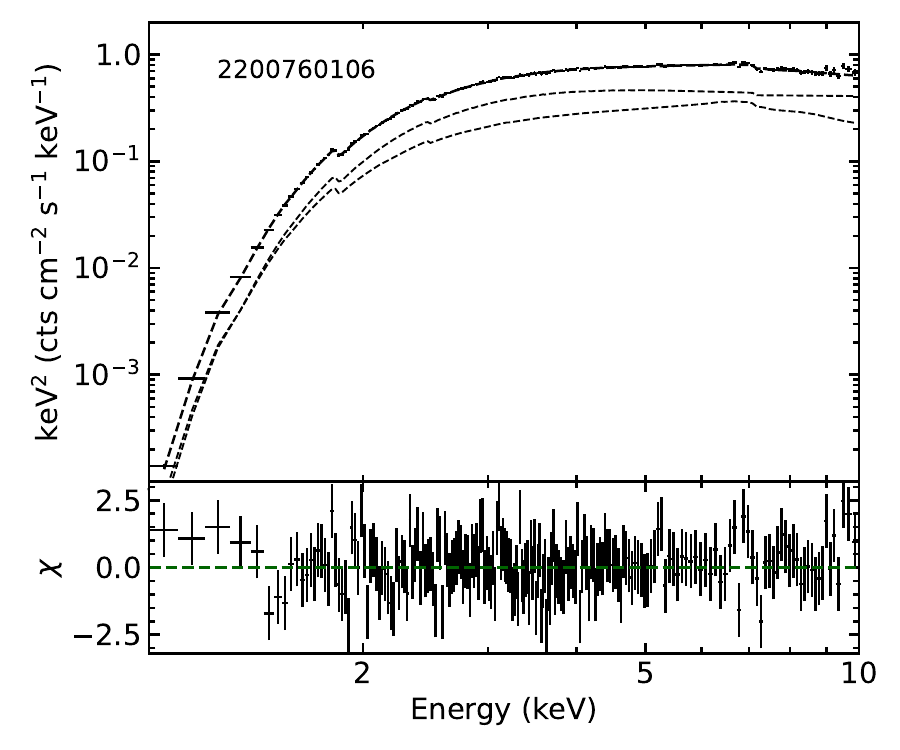} \\ 
    \end{center}
    \vspace{-0.2cm}
    \caption{Left: a typical PDS of \srce\ from \nicer\ data (Obs ID: 2200760106). Right: the \nicer\ spectrum and residuals to the best-fit model for the same observation. 
    \label{nicer_exo1846}}
\end{figure*}

The left panels of Figs.~\ref{hxmt_gx339}--\ref{nicer_exo1846} show typical PDS and fit residuals for HXMT data and \nicer\ data of \srcg\ and \srce, respectively, and the right panels show typical spectra and fit residuals. In Tabs.~\ref{gx339_obs}--\ref{exo1846_obs}, we summarize the selected observations that are analyzed in this work. Tabs.~\ref{result_gx339}--\ref{result_exo1846} show the best-fit results of \srcg\ and \srce, respectively.


\begin{table*}[]
\centering
\vspace{0.5cm}
\begin{tabular}{cccc}
\hline\hline
Mission \hspace{0.2cm} & \hspace{0.2cm} Obs.~ ID  \hspace{0.1cm} & \hspace{0.1cm}  Start data \hspace{0.1cm} & \hspace{0.1cm}  Exposure (s) \hspace{0.1cm}  \\ \hline
\multirow{6}*{\hxmt}  & P0304024026 &  2021-03-12 & 2274 \\ 
                      & P0304024028 & 2021-03-14 & 1401 \\ 
                      & P0304024032 & 2021-03-18 & 1597 \\ 
                      & P0304024035 & 2021-03-22 & 1669 \\
                      & P0304024036 & 2021-03-24 & 1193 \\
                      & P0304024038 & 2021-03-26 & 2088 \\ 
            \hline
\multirow{6}*{\nicer} & 3558011402  &  2021-03-17 & 1595 \\
                       & 3558011501  & 2021-03-19 & 7560 \\
                       & 4133010101  & 2021-03-19 & 2030 \\
                       & 4133010102  & 2021-03-20 & 1860 \\
                       & 4133010103  & 2021-03-26 & 6111 \\
                       & 4133010104  & 2021-03-27 & 8709 \\
\hline\hline
\end{tabular}
\vspace{0.3cm}
\caption{\rm \hxmt\ and \nicer\ observations of \srcg\ analyzed in this work. For \hxmt, the listed exposure time is for the LE instrument.
\label{gx339_obs}}
\end{table*}

\begin{table*}[]
\centering
\vspace{0.5cm}
\begin{tabular}{cccc}
\hline\hline
Mission \hspace{0.2cm} & \hspace{0.2cm} Obs.~ ID  \hspace{0.1cm} & \hspace{0.1cm}  Start data \hspace{0.1cm} & \hspace{0.1cm}  Exposure (s) \hspace{0.1cm}  \\ \hline
\multirow{16}*{\hxmt} & P021405000101 & 2019-08-02 & 718 \\
                      & P021405000102  & 2019-08-02 & 1436 \\
                      & P021405000103  & 2019-08-02 & 762 \\
                      & P021405000104  & 2019-08-02 & 718 \\
                      & P021405000105  & 2019-08-02 & 1715 \\
                      & P021405000106  & 2019-08-03 & 563 \\
                      & P021405000107  & 2019-08-03 & 656 \\
                      & P021405000301  & 2019-08-05 & 700 \\
                      & P021405000302  & 2019-08-05 & 1102 \\
                      & P021405000303  & 2019-08-05 & 678 \\
                      & P021405000401  & 2019-08-06 & 718 \\
                      & P021405000502  & 2019-08-07 & 691 \\
                      & P021405000503  & 2019-08-07 & 539 \\
                      & P021405000601  & 2019-08-08 & 1130 \\
                      & P021405000701  & 2019-08-08 & 1163 \\
                      & P021405000702  & 2019-08-09 & 1795 \\
            \hline 
\multirow{19}*{\nicer} & 2200760101 & 2019-07-31 & 5658 \\
                       &2200760102 & 2019-08-01 & 1165 \\
                       &2200760103 & 2019-08-02 & 2562 \\
                       &2200760104 & 2019-08-03 & 1488 \\
                       &2200760105 & 2019-08-04 & 1130 \\
                       &2200760106 & 2019-08-05 & 3564 \\
                       &2200760107 & 2019-08-06 & 912 \\
                       &2200760108 & 2019-08-07 & 927 \\
                       &2200760109 & 2019-08-08 & 3293 \\
                       &2200760110 & 2019-08-09 & 4629 \\
                       &2200760112 & 2019-08-11 & 2749 \\
                       &2200760113 & 2019-08-12 & 3341 \\
                       &2200760114 & 2019-08-13 & 7154 \\
                       &2200760115 & 2019-08-13 & 8181 \\
                       &2200760116 & 2019-08-15 & 4703 \\
                       &2200760117 & 2019-08-16 & 8739 \\
                       &2200760118 & 2019-08-17 & 4875 \\
                       &2200760119 & 2019-08-17 & 3341 \\
                       &2200760120 & 2019-08-19 & 3894 \\
\hline\hline
\end{tabular}
\vspace{0.3cm}
\caption{\rm \hxmt\ and \nicer\ observations of \srce\ analyzed in this work. As in the case of \srcg, the listed exposure time is for the LE instrument of \hxmt. 
\label{exo1846_obs}}
\end{table*}


\begin{table*}[]
\centering
\vspace{0.5cm}
\hspace{-1.0cm}
\setlength{\tabcolsep}{2.05pt}
\begin{tabular}{cccccccccc}
\hline\hline
Obs.~ID & $N_{\rm H}$ [$10^{22}$~cm$^{-2}$] & $f_{\rm QPO}$ & $ \Gamma $ &$ f_{\rm sc} $ &$\dot{M}$  [$10^{15}$~g~s$^{-1}$]  &$ R_{\rm in}$ [$r_{\rm g}$] & $ \log F{\rm con}$ & $ \log F{\rm ref}$ & $\chi^2/\nu $  \\  \hline
P0304024026 &$0.55^{*}$ & $0.276_{-0.025}^{+0.022}$ & $1.748_{-0.025}^{+0.022}$ & $0.40_{-0.21}^{+P}$ & $0.39_{-0.13}^{+1.7}$ & $40_{-13}^{+57}$ & $-8.38_{-0.25}^{+0.26}$ & $-8.76_{-0.25}^{+0.21}$ & $893.93/1057$ \\
P0304024028 &$0.55^{*}$ & $0.370_{-0.012}^{+0.016}$ & $1.788_{-0.03}^{+0.023}$ & $0.6_{-0.4}^{+0.17}$ & $0.21_{-0.05}^{+1.2}$ & $22_{-6}^{+13}$ & $-8.34_{-0.19}^{+0.06}$ & $-8.89_{-0.4}^{+0.20}$ & $833.81/959$  \\
P0304024032 &$0.55^{*}$ & $0.428_{-0.019}^{+0.013}$ & $1.812_{-0.014}^{+0.025}$ & $0.44_{-0.11}^{+P}$ & $0.8_{-0.4}^{+2.9}$ & $55.6_{-7}^{+44}$ & $-8.17_{-0.3}^{+0.18}$ & $-8.92_{-0.18}^{+0.3}$ & $911.31/1013$ \\
P0304024035 &$0.55^{*}$ & $0.66\pm0.04$ & $1.851_{-0.024}^{+0.028}$ & $0.56_{-0.11}^{+0.09}$ & $0.31_{-0.08}^{+0.3}$ & $18.2_{-2.4}^{+13}$ & $-8.16_{-0.15}^{+0.06}$ & $-8.89_{-0.25}^{+0.4}$ & $930.44/1033$ \\  
P0304024036 &$0.55^{*}$ & $0.93\pm0.05$ & $1.831_{-0.04}^{+0.022}$ & $0.50\pm0.04$ & $0.33\pm0.05$ & $15.5_{-2.7}^{+3.0}$ & $-8.12_{-0.3}^{+0.25}$ & $-8.52_{-0.24}^{+0.3}$ & $948.06/1021$ \\
P0304024038 &$0.55^{*}$ & $2.53\pm0.06$ & $2.338_{-0.023}^{+0.021}$ & $0.432_{-0.017}^{+0.016}$ & $0.84_{-0.04}^{+0.03}$ & $15.4_{-0.7}^{+0.6}$ & $-8.01_{-0.5}^{+0.3}$ & $-8.61_{-0.15}^{+0.23}$ & $959.94/977$ \\  
\hline
3558011402 &$0.62\pm0.04$ & $0.407\pm0.019$ & $1.703_{-0.024}^{+0.023}$ & $0.5_{-0.4}^{+0.20}$ & $0.35_{-0.18}^{+0.20}$ & $36.3_{-12}^{+8}$ & $-8.30_{-0.7}^{+0.07}$ & $-8.60_{-0.11}^{+0.2}$ & $684.37/852$ \\
3558011501 &$0.557\pm0.004$ & $0.492_{-0.015}^{+0.018}$ & $1.7186_{-0.0024}^{+0.0022}$ & $0.69_{-0.10}^{+0.11}$ & $0.207\pm0.005$ & $21.7\pm1.1$ & $-8.303\pm0.017$ & $-8.67\pm0.04$ & $587.78/899$ \\
4133010101 & $0.617_{-0.025}^{+0.12}$ & $0.491_{-0.017}^{+0.03}$ & $1.711_{-0.029}^{+0.02}$ & $0.48_{-0.22}^{+0.21}$ & $0.30_{-0.09}^{+0.16}$ & $31.1_{-7}^{+10}$ & $-8.38_{-0.29}^{+0.06}$ & $-8.48_{-0.09}^{+0.11}$ & $725.14/889$ \\
4133010102 & $0.580_{-0.018}^{+0.04}$ & $0.520_{-0.028}^{+0.026}$ & $1.715_{-0.024}^{+0.019}$ & $0.6_{-0.5}^{+0.20}$ & $0.23_{-0.06}^{+0.04}$ & $23.1_{-1.9}^{+4}$ & $-8.32_{-0.6}^{+0.06}$ & $-8.61_{-0.06}^{+0.16}$ & $691.04/886$ \\
4133010103-1 & $0.457_{-0.03}^{+0.018}$ & $1.97_{-0.06}^{+0.05}$ & $2.247_{-0.018}^{+0.03}$ & $0.517_{-0.04}^{+0.005}$ & $0.64_{-0.11}^{+0.16}$ & $20\pm3$ & $-8.27_{-0.12}^{+0.09}$ & $-8.67_{-0.3}^{+0.20}$ & $581.61/709$ \\
4133010103-2 & $0.487_{-0.029}^{+0.021}$ & $1.98_{-0.06}^{+0.04}$ & $2.12_{-0.06}^{+0.05}$ & $0.32_{-0.3}^{+0.11}$ & $0.57\pm0.25$ & $18\pm5$ & $-8.27_{-0.6}^{+0.15}$ & $-8.28_{-0.23}^{+0.22}$ & $604.52/719$ \\
4133010103-3 & $0.46\pm0.03$ & $2.20\pm0.04$ & $2.16_{-0.06}^{+0.05}$ & $0.43_{-0.19}^{+0.07}$ & $0.41_{-0.11}^{+0.08}$ & $13.9_{-2.2}^{+2.1}$ & $-8.28_{-0.24}^{+0.09}$ & $-8.66_{-0.28}^{+0.4}$ & $594.45/683$ \\
4133010103-4 & $0.489\pm0.027$ &$2.54_{-0.04}^{+0.05}$ & $2.26_{-0.07}^{+0.08}$ & $0.38_{-0.25}^{+0.10}$ & $0.39_{-0.13}^{+0.07}$ & $12\pm3$ & $-8.29_{-0.4}^{+0.11}$ & $-8.51_{-0.27}^{+0.3}$ & $609.25/715$ \\
4133010103-5 & $0.481_{-0.023}^{+0.027}$ & $2.69_{-0.05}^{+0.04}$ & $2.30_{-0.12}^{+0.05}$ & $0.45_{-0.15}^{+0.07}$ & $0.46_{-0.13}^{+0.11}$ & $12.9_{-2.7}^{+2.6}$ & $-8.20_{-0.14}^{+0.06}$ & $-8.68\pm0.26$ & $605.14/718$ \\
4133010103-6 & $0.473_{-0.018}^{+0.021}$ & $2.55_{-0.04}^{+0.03}$ & $2.20_{-0.07}^{+0.06}$ & $0.37_{-0.13}^{+0.07}$ & $0.46_{-0.12}^{+0.11}$ & $12.7\pm2.5$ & $-8.21_{-0.18}^{+0.06}$ & $-8.49_{-0.14}^{+0.22}$ & $550.91/763$ \\
4133010103-7 & $0.488_{-0.022}^{+0.023}$ & $2.64_{-0.02}^{+0.03}$ & $2.28_{-0.09}^{+0.06}$ & $0.39_{-0.13}^{+0.08}$ & $0.47_{-0.11}^{+0.13}$ & $13.2_{-2.6}^{+2.8}$ & $-8.23_{-0.13}^{+0.09}$ & $-8.43_{-0.19}^{+0.15}$ & $641.36/781$ \\
4133010103-8 & $0.495_{-0.021}^{+0.018}$ & $2.91_{-0.04}^{+0.03}$ & $2.35_{-0.08}^{+0.06}$ & $0.44_{-0.11}^{+0.08}$ & $0.51_{-0.11}^{+0.13}$ & $12.7_{-2.4}^{+2.7}$ & $-8.17_{-0.11}^{+0.06}$ & $-8.50\pm0.17$ & $609.30/776$ \\
4133010103-9 & $0.504_{-0.02}^{+0.015}$ & $3.06_{-0.02}^{+0.03}$ & $2.35_{-0.11}^{+0.06}$ & $0.40_{-0.14}^{+0.09}$ & $0.44_{-0.09}^{+0.11}$ & $11.0_{-2.3}^{+2.2}$ & $-8.23_{-0.13}^{+0.08}$ & $-8.35\pm0.13$ & $620.33/788$ \\
4133010104-1 & $0.500_{-0.019}^{+0.020}$ & $3.28\pm0.05$ & $2.35_{-0.16}^{+0.09}$ & $0.37_{-0.22}^{+0.16}$ & $0.42_{-0.13}^{+0.14}$ & $10.1_{-2.6}^{+2.7}$ & $-8.22_{-0.5}^{+0.10}$ & $-8.39_{-0.22}^{+0.29}$ & $645.09/762$ \\ 
4133010104-2 & $0.502_{-0.021}^{+0.023}$ & $3.20_{-0.04}^{+0.03}$ & $2.39_{-0.09}^{+0.05}$ & $0.45_{-0.10}^{+0.06}$ & $0.51_{-0.12}^{+0.08}$ & $11.9_{-2.5}^{+1.5}$ & $-8.15_{-0.11}^{+0.05}$ & $-8.47_{-0.14}^{+0.17}$ & $618.63/763$ \\ 
4133010104-3 & $0.500_{-0.029}^{+0.027}$ & $3.04_{-0.26}^{+0.14}$ & $2.46_{-0.22}^{+0.07}$ & $0.50_{-0.29}^{+0.08}$ & $0.69_{-0.17}^{+0.19}$ & $13.2_{-4}^{+2.8}$ & $-8.04_{-0.5}^{+0.07}$ & $-8.46_{-0.28}^{+0.4}$ & $547.62/656$ \\ 
4133010104-4 & $0.507_{-0.029}^{+0.028}$ & $3.04_{-0.26}^{+0.14}$ & $2.34_{-0.22}^{+0.14}$ & $0.36_{-P}^{+0.21}$ & $0.6\pm0.3$ & $12\pm5$ & $-8.11_{-0.5}^{+0.11}$ & $-8.3_{-0.8}^{+0.3}$ & $565.97/656$ \\
4133010104-5 & $0.48\pm0.03$ & $3.41_{-0.06}^{+0.07}$ & $2.4_{-0.3}^{+0.10}$ & $0.41_{-P}^{+0.19}$ & $0.34_{-0.17}^{+0.08}$ & $8.5_{-2.8}^{+2.9}$ & $-8.26_{-0.5}^{+0.07}$ & $-8.6_{-0.6}^{+0.4}$ & $502.43/616$ \\ 
4133010104-6 & $0.50\pm0.02$ & $3.63\pm0.06$ & $2.38_{-0.12}^{+0.09}$ & $0.40_{-0.14}^{+0.09}$ & $0.50_{-0.11}^{+0.12}$ & $10.6\pm2.2$ & $-8.14_{-0.12}^{+0.07}$ & $-8.50_{-0.21}^{+0.20}$ & $559.10/776$ \\ 
4133010104-7 & $0.500_{-0.022}^{+0.019}$ & $3.97_{-0.08}^{+0.10}$ & $2.39_{-0.18}^{+0.11}$ & $0.37_{-0.17}^{+0.1}$ & $0.36_{-0.08}^{+0.09}$ & $8.2_{-1.9}^{+1.8}$ & $-8.23_{-0.15}^{+0.07}$ & $-8.56\pm0.22$ & $559.05/700$  \\ 
4133010104-8 & $0.534_{-0.024}^{+0.03}$ & $3.99_{-0.06}^{+0.05}$ & $2.52_{-0.28}^{+0.17}$ & $0.40_{-0.4}^{+0.23}$ & $0.34_{-0.14}^{+0.11}$ & $7.7_{-2.9}^{+2.8}$ & $-8.26_{-0.4}^{+0.12}$ & $-8.46_{-0.23}^{+0.3}$ & $515.87/692$ \\ 
4133010104-9 & $0.512_{-0.026}^{+0.023}$ & $3.99_{-0.06}^{+0.05}$ & $2.50_{-0.19}^{+0.13}$ & $0.37_{-0.18}^{+0.1}$ & $0.34_{-0.08}^{+0.07}$ & $7.1_{-1.3}^{+1.4}$ & $-8.25_{-0.14}^{+0.07}$ & $-8.53_{-0.20}^{+0.19}$ & $611.50/685$\\ 
4133010104-10 & $0.528_{-0.018}^{+0.019}$ & $4.55\pm0.08$ & $2.57_{-0.15}^{+0.11}$ & $0.36_{-0.13}^{+0.1}$ & $0.48\pm0.09$ & $7.4_{-1.1}^{+1.5}$ & $-8.10_{-0.12}^{+0.05}$ & $-8.44_{-0.16}^{+0.19}$ & $591.18/774$ \\ 
4133010104-11 & $0.500_{-0.018}^{+0.022}$ & $4.34\pm0.04$ & $2.42_{-0.21}^{+0.14}$ & $0.30_{-0.17}^{+0.12}$ & $0.38_{-0.07}^{+0.06}$ & $6.8_{-1.3}^{+1.4}$ & $-8.19_{-0.10}^{+0.07}$ & $-8.51_{-0.20}^{+0.22}$ & $544.57/701$ \\
4133010104-12 & $0.526_{-0.022}^{+0.025}$ & $4.81\pm0.07$ & $2.62_{-0.20}^{+0.13}$ & $0.37_{-0.13}^{+0.11}$ & $0.42_{-0.09}^{+0.08}$ & $6.8\pm1.4$ & $-8.14_{-0.14}^{+0.05}$ & $-8.60_{-0.18}^{+0.26}$ & $536.68/698$ \\
4133010104-13 & $0.528_{-0.024}^{+0.016}$ & $5.37_{-0.19}^{+0.23}$ & $2.70_{-0.29}^{+0.14}$ & $0.39_{-0.20}^{+0.12}$ & $0.54_{-0.10}^{+0.09}$ & $7.7\pm1.7$ & $-8.06_{-0.18}^{+0.04}$ & $-8.67_{-0.24}^{+0.4}$ & $564.02/718$ \\
\hline\hline
\end{tabular}
\vspace{0.3cm}
\caption{\rm Best-fit Parameters of \srcg. The symbol $*$ indicates that the parameter is frozen in the fit. $\log F_{\rm con}$ and $\log F_{\rm ref}$ represent the 1–20~keV flux of continuum (disk and Comptonisation) and reflection components, respectively. The symbol $P$ means that the error bar touches the lower (or higher) limit. All uncertainties are quoted at the 90\% confidence level. 
\label{result_gx339}}
\end{table*}



\begin{table*}[]
\centering
\vspace{0.5cm}
\hspace{-1.0cm}
\setlength{\tabcolsep}{2.05pt}
\begin{tabular}{ccccccccccc}
\hline\hline
 Obs.~ID & $N_{\rm H}$ [$10^{22}$~cm$^{-2}$] & QPO frequency & $ \Gamma $ &$ f_{\rm sc} $ &$\dot{M}$  [$10^{15}$~g~s$^{-1}$]  &$ R_{\rm in}$ [$r_{\rm g}$] & $ \log F{\rm con}$ & $ \log F{\rm ref}$ & $\chi^2/\nu $  \\  \hline
P021405000101 & $6.73_{-0.19}^{+0.24}$ & $0.689_{-0.009}^{+0.011}$ & $1.81_{-0.04}^{+0.07}$ & $0.09_{-0.04}^{+0.08}$ & $6.0_{-2.2}^{+1.1}$ & $99_{-58}^{+P}$ & $-8.04_{-0.16}^{+0.11}$ & $-8.74_{-0.24}^{+0.5}$ & $1016.51/1051$ \\
P021405000102 & $7.4_{-0.7}^{+0.8}$ & $0.851_{-0.012}^{+0.011}$ & $1.82\pm0.04$ & $0.32_{-0.11}^{+0.21}$ & $0.9_{-0.5}^{+0.9}$ & $25_{-10}^{+20}$ & $-8.05\pm0.09$ & $-8.75_{-0.21}^{+0.5}$ & $988.95/1022$ \\
P021405000103 & $8.7_{-1.1}^{+1.5}$ & $0.994\pm0.012$ & $1.95_{-0.05}^{+0.04}$ & $0.13_{-0.03}^{+0.05}$ & $5.2_{-1.8}^{+2.1}$ & $55_{-16}^{+41}$ & $-7.92_{-0.18}^{+0.09}$ & $-8.77_{-0.28}^{+0.5}$ & $1026.72/1022$ \\
P021405000104 & $7.6_{-0.5}^{+0.6}$ & $1.132_{-0.021}^{+0.019}$ & $2.02_{-0.06}^{+0.06}$ & $0.17_{-0.05}^{+0.16}$ & $6.4_{-4}^{+2.5}$ & $83_{-50}^{+17}$ & $-8.08_{-0.11}^{+0.09}$ & $-8.50_{-0.25}^{+0.08}$ & $1030.04/1019$ \\
P021405000105 & $8.0_{-0.5}^{+P}$ & $1.251\pm0.014$ & $1.964_{-0.024}^{+0.05}$ & $0.20_{-0.03}^{+0.10}$ & $2.8_{-1.3}^{+2.0}$ & $41_{-12}^{+15}$ & $-8.03_{-0.05}^{+0.03}$ & $-8.5\pm0.4$ & $1073.18/1139$ \\
P021405000106 & $6.13_{-0.17}^{+0.3}$ & $1.340_{-0.011}^{+0.009}$ & $2.029_{-0.017}^{+0.06}$ & $0.989_{0.021}^{+P}$ & $0.29_{-0.04}^{+0.09}$ & $9.94_{-0.16}^{+4}$ & $-8.18_{-0.3}^{+0.04}$ & $-8.51_{-0.25}^{+0.06}$ & $994.17/922$ \\
P021405000107 & $6.59_{-0.5}^{+0.27}$ & $1.60_{-0.05}^{+0.04}$ & $2.05\pm0.07$ & $0.51_{-0.29}^{+P}$ & $1.0_{-0.3}^{+0.9}$ & $29_{-8}^{+20}$ & $-8.03_{-1.9}^{+0.15}$ & $-8.8_{-1.0}^{+0.8}$ & $954.56/967$ \\
P021405000201 & $6.35_{-0.23}^{+0.5}$ & $2.26_{-0.06}^{+0.04}$ & $2.245_{-0.028}^{+0.05}$ & $0.28_{-0.08}^{+0.4}$ & $6_{-4}^{+6}$ & $100_{-56}^{+P}$ & $-8.00_{-0.6}^{+0.12}$ & $-8.5\pm0.4$ & $947.89/1006$ \\
P021405000301 & $5.5_{-0.5}^{+0.4}$ & $3.14\pm0.03$ & $2.41_{-0.07}^{+0.09}$ & $0.55_{-0.06}^{+0.12}$ & $0.39_{-0.15}^{+0.3}$ & $6.0_{-2.3}^{+6}$ & $-8.00_{-0.15}^{+0.03}$ & $-8.8_{-0.5}^{+0.6}$ & $1043.07/1056$ \\
P021405000302 & $6.12_{-0.26}^{+0.4}$ & $3.18\pm0.07$ & $2.43\pm0.07$ & $0.50\pm0.05$ & $0.36_{-0.10}^{+0.15}$ & $5.5_{-1.2}^{+3}$ & $-8.14_{-0.15}^{+0.16}$ & $-8.1_{-0.7}^{+0.14}$ & $1007.96/1113$ \\
P021405000303 & $6.2_{-0.7}^{+1.2}$ & $3.73\pm0.05$ & $2.11_{-0.11}^{+0.18}$ & $0.004_{-P}^{+0.5}$ & $0.09\pm0.06$ & $3.5\pm1.2$ & $-8.64_{-0.08}^{+0.8}$ & $-7.89_{-0.8}^{+0.04}$ & $889.98/912$ \\
P021405000401 & $5.1\pm0.6$ & $3.26_{-0.11}^{+0.13}$ & $2.22_{-0.06}^{+0.10}$ & $0.47_{-0.06}^{+0.09}$ & $0.29_{-0.09}^{+0.19}$ & $4.4_{-1.4}^{+2.5}$ & $-7.93_{-0.4}^{+0.05}$ & $-8.9_{-1.1}^{+0.9}$ & $835.45/909$ \\
P021405000502 & $5.6_{-0.6}^{+0.3}$ & $4.19_{-0.13}^{+0.15}$ & $2.43_{-0.04}^{+0.09}$ & $0.420_{-0.05}^{+0.018}$ & $0.50_{-0.17}^{+0.18}$ & $6.0_{-1.6}^{+0.7}$ & $-7.90_{-0.04}^{+0.03}$ & $-9.1_{-0.3}^{+0.6}$ & $993.15/989$ \\
P021405000503 & $6.4_{-0.5}^{+0.7}$ & $3.90\pm0.09$ & $2.54_{-0.11}^{+0.1}$ & $0.35_{-0.05}^{+0.04}$ & $0.62_{-0.24}^{+0.17}$ & $6.8_{-2.2}^{+1.1}$ & $-8.02_{-0.06}^{+0.09}$ & $-8.21_{-0.3}^{+0.15}$ & $937.30/953$ \\
P021405000601 & $6.01_{-0.4}^{+0.27}$ & $4.48\pm0.05$ & $2.57_{-0.08}^{+0.06}$ & $0.403_{-0.03}^{+0.027}$ & $0.75_{-0.19}^{+0.12}$ & $7.4_{-1.6}^{+0.9}$ & $-7.92_{-0.05}^{+0.08}$ & $-8.29_{-0.28}^{+0.19}$ & $1047.54/1021$ \\
P021405000701 & $6.9_{-0.5}^{+0.3}$ & $6.69\pm0.33$ & $2.44_{-0.09}^{+0.05}$ & $0.123_{--0.067}^{+0.07}$ & $0.84_{-0.25}^{+0.4}$ & $7.1_{-1.1}^{+1.0}$ & $-7.98_{-0.12}^{+0.26}$ & $-7.83_{-0.23}^{+0.10}$ & $1059.74/1136$ \\
P021405000702 & $6.24_{-0.29}^{+0.3}$ & $6.31_{-0.13}^{+0.14}$ & $2.42_{-0.06}^{+0.10}$ & $0.209_{-0.022}^{+0.025}$ & $1.16_{-0.13}^{+0.14}$ & $7.4\pm0.7$ & $-7.69_{-0.21}^{+0.03}$ & $-9.01_{-0.18}^{+0.4}$ & $1062.41/1101$ \\
\hline
2200760101 & $6.60_{-0.24}^{+0.23}$ & $0.262_{-0.012}^{+0.008}$ & $1.609_{-0.04}^{+0.017}$ & $0.16_{-0.06}^{+0.15}$ & $0.57_{-0.08}^{+0.27}$ & $33_{-14}^{+25}$ & $-8.45_{-0.07}^{+0.08}$ & $-8.84_{-0.20}^{+0.08}$ & $799.14/899$ \\
2200760102 & $7.2_{-0.3}^{+0.4}$ & $0.412_{-0.015}^{+0.017}$ & $1.81_{-0.06}^{+0.11}$ & $0.053_{-0.007}^{+0.016}$ & $3.1\pm1.0$ & $90_{-28}^{+6}$ & $-8.70_{-0.03}^{+0.14}$ & $-9.0\pm0.4$ & $704.88/672$ \\
2200760103 & $6.83_{-0.2}^{+0.12}$ & $1.007_{-0.028}^{+0.029}$ & $1.972_{-0.026}^{+0.04}$ & $0.257_{-0.04}^{+0.026}$ & $0.71_{-0.11}^{+0.20}$ & $33_{-3}^{+4}$ & $-8.49_{-0.15}^{+0.12}$ & $-8.85_{-0.18}^{+0.12}$ & $861.35/853$ \\
2200760104 & $6.9_{-0.4}^{+0.3}$ & $1.37_{-0.06}^{+0.05}$  & $1.877_{-0.05}^{+0.027}$ & $0.10_{-P}^{+0.23}$ & $0.9_{-0.6}^{+0.7}$ & $32_{-9}^{+11}$ & $-8.5_{-0.3}^{+0.5}$ & $-8.6_{-0.03}^{+0.18}$ & $841.57/849$ \\
2200760105 & $5.66_{-0.11}^{+0.11}$ & $2.43_{-0.06}^{+0.06}$ & $1.85_{-0.11}^{+0.12}$ & $0.39_{-0.07}^{+0.05}$ & $0.020\pm0.004$ & $1.9\pm0.4$ & $-8.591_{-0.010}^{+0.012}$ & $-8.90_{-0.03}^{+0.05}$ & $728.60/759$ \\
2200760106 & $5.90_{-0.08}^{+0.09}$ & $3.25\pm0.04$ & $2.15_{-0.11}^{+0.16}$ & $0.46_{-0.13}^{+0.17}$ & $0.04_{-0.02}^{+0.03}$ & $2.8_{-1.3}^{+1.2}$ & $-8.61_{-0.19}^{+0.16}$ & $-8.76_{-0.23}^{+0.19}$ & $780.47/863$ \\
2200760107 & $6.20_{-0.10}^{+0.11}$ & $3.31\pm0.05$ & $2.367_{-0.028}^{+0.03}$ & $0.6^{*}$ & $0.07_{-0.03}^{+0.07}$ & $4.9_{-1.6}^{+2.9}$ & $-8.49\pm0.14$ & $-8.33_{-0.15}^{+0.09}$ & $755.51/798$ \\
2200760108 & $5.99_{-0.12}^{+0.17}$ & $3.69\pm0.06$ & $2.50_{-0.04}^{+0.07}$ & $0.6^{*}$ & $0.179_{-0.07}^{+0.026}$ & $6.3_{-1.7}^{+0.7}$ & $-8.24_{-0.07}^{+0.04}$ & $-8.66_{-0.16}^{+0.19}$ & $709.60/798$ \\
2200760109 & $6.32_{-0.11}^{+0.09}$ & $6.93_{-0.3 }^{+0.28}$ & $2.20_{-0.6}^{+0.18}$ & $0.12_{-P}^{+0.11}$ & $0.87_{-0.11}^{+0.07}$ & $6.3\pm0.6$ & $-7.81_{-0.16}^{+0.05}$ & $-8.16_{-0.07}^{+0.23}$ & $755.68/898$ \\
2200760110 & $6.19_{-0.04}^{+0.09}$ & $6.46_{-0.28}^{+0.5 }$ & $1.80_{-0.28}^{+0.6}$ & $0.04_{-0.04}^{+0.11}$ & $0.95_{-0.10}^{+0.06}$ & $6.3_{-0.4}^{+0.6}$ & $-7.81_{-0.10}^{+0.06}$ & $-8.15_{-0.18}^{+0.14}$ & $750.97/899$ \\
2200760112 & $6.35_{-0.06}^{+0.08}$ & $6.16_{-0.34}^{+0.5}$ & $2.40_{-0.23}^{+0.12}$ & $0.22_{-0.06}^{+0.05}$ & $0.86_{-0.10}^{+0.08}$ & $7.3_{-0.4}^{+0.6}$ & $-7.83_{-0.05}^{+0.08}$ & $-8.23_{-0.06}^{+0.13}$ & $784.13/889$ \\
2200760113 & $6.56_{-0.15}^{+0.11}$ & $6.6_{-0.5}^{+1.2}$ & $2.50_{-0.24}^{+0.14}$ & $0.17_{-P}^{+0.07}$ & $0.12_{-0.05}^{+0.04}$ & $3.9_{-0.7}^{+0.5}$ & $-8.47_{-0.21}^{+0.09}$ & $-8.59_{-0.23}^{+0.17}$ & $782.20/858$  \\
2200760114 & $6.28_{-0.03}^{+0.15}$ & $5.93_{-0.20}^{+0.5}$ & $2.17_{-0.17}^{+0.27}$ & $0.08_{-P}^{+0.11}$ & $0.40_{-0.08}^{+0.06}$ & $5.6_{-0.5}^{+0.6}$ & $-8.13_{-0.14}^{+0.04}$ & $-8.25_{-0.3}^{+0.08}$ & $766.57/899$ \\
2200760115 & $6.35_{-0.14}^{+0.11}$ & $5.76_{-0.23}^{+0.26}$ & $2.33_{-0.19}^{+0.4}$ & $0.12_{-P}^{+0.25}$ & $0.13_{-0.06}^{+0.04}$ & $4.0_{-0.5}^{+1.1}$ & $-8.45_{-0.14}^{+0.18}$ & $-8.36_{-0.10}^{+0.10}$ & $715.46/899$ \\
2200760116 & $6.19_{-0.1}^{+0.12}$ & $5.35_{-0.18}^{+0.16}$ & $2.27_{-0.22}^{+0.4}$ & $0.18_{-0.11}^{+0.3}$ & $0.23_{-0.06}^{+0.05}$ & $5.2_{-0.5}^{+2.8}$ & $-8.29_{-0.11}^{+0.08}$ & $-8.38_{-0.4}^{+0.12}$ & $812.60/899$ \\
2200760117 & $6.31_{-0.14}^{+0.12}$ & $5.27_{-0.24}^{+0.24}$ & $2.25_{-0.23}^{+0.5}$ & $0.09_{-P}^{+0.3}$ & $0.21_{-0.06}^{+0.07}$ & $5.0_{-0.5}^{+2.5}$ & $-8.38_{-0.14}^{+0.22}$ & $-8.36_{-0.21}^{+0.20}$ & $723.59/899$ \\
2200760118 & $6.54_{-0.09}^{+0.11}$ & $5.77_{-0.23}^{+0.26}$ & $2.81_{-0.13}^{+0.11}$ & $0.50_{-0.10}^{+0.15}$ & $0.40_{-0.07}^{+0.12}$ & $7.8_{-1.8}^{+2.4}$ & $-8.14_{-0.07}^{+0.05}$ & $-8.49_{-0.13}^{+0.06}$ & $766.21/899$ \\
2200760119 & $6.13_{-0.08}^{+0.05}$ & $5.18_{-0.12}^{+0.15}$ & $2.70_{-0.18}^{+0.4}$ & $0.13_{-0.06}^{+0.11}$ & $0.281_{-0.029}^{+0.04}$ & $4.9_{-0.4}^{+0.7}$ & $-8.30_{-0.07}^{+0.06}$ & $-8.71_{-0.18}^{+0.23}$ & $755.06/856$ \\
2200760120 & $6.41_{-0.3}^{+0.26}$ & $5.34_{-0.15}^{+0.22}$ & $2.96_{-0.4}^{+0.28}$ & $0.22_{-0.07}^{+0.14}$ & $0.148_{-0.029}^{+0.027}$ & $4.2_{-0.7}^{+0.5}$ & $-8.47_{-0.19}^{+0.09}$ & $-8.82_{-0.16}^{+0.13}$ & $736.47/837$ \\
\hline\hline
\end{tabular}
\vspace{0.3cm}
\caption{\rm Best-fit Parameters of \srce. It is organized as in Tab.~\ref{result_gx339}. All uncertainties are quoted at the 90\% confidence level. 
\label{result_exo1846}}
\end{table*}


\bibliography{QPOs_GX339}{}
\bibliographystyle{aasjournal}

\end{document}